\documentclass{aa}  
\usepackage{graphicx}
\usepackage{txfonts}
\usepackage{stfloats}
\usepackage{tabularx}
\usepackage[table,xcdraw]{xcolor}
\usepackage{url}
\usepackage{xcolor}
\usepackage{natbib}
\bibpunct{(}{)}{;}{a}{}{,} 
\usepackage{array}
\usepackage[colorlinks=true, linkcolor=blue, citecolor=blue, urlcolor=blue]{hyperref}
\usepackage{gensymb}
\usepackage{subfigure}
\usepackage{amsmath}
\usepackage{orcidlink}
\usepackage{comment}
\usepackage{svg}
\usepackage{placeins} 
\usepackage{subfigure}
\usepackage{xspace}

\newcommand{\kcl}[0]{\text{KCl} }

\newcommand{\nas}{\ensuremath{\mathrm{Na_2S}}\xspace}
\newcommand{\mg}{\ensuremath{\mathrm{MgSiO_3}}\xspace}

\newcommand{\micron}{\ensuremath{\mu\mathrm{m}}\xspace}

\begin{document} 

\title{The tale of the 3 planets: 3D cloud feedback enhances the spectral diversity of warm Jupiters}
\author{Nishil Mehta\inst{1}\thanks{nishil.mehta@oca.eu}
       \and
       Vivien Parmentier\inst{1}
       \and
       Xianyu Tan\inst{2}
       \and
       Elspeth K. H. Lee\inst{3}
       \and
       Tristan Guillot\inst{1}
       \and
       Matthew M. Murphy\inst{4}
       \and
       Thomas P. Greene\inst{5}
       \and
       Thomas G. Beatty\inst{6}
       \and
       Taylor J. Bell\inst{7}
       \and
       Jonathan J. Fortney\inst{8}
       \and
       Michael R. Line\inst{9}
       \and
       Sagnick Mukherjee\inst{9}
       \and
       Kazumasa Ohno\inst{10}
       \and
       Everett Schlawin\inst{11}
       \and
       Anastasia Triantafillides\inst{6}
       \and
       Luis Welbanks\inst{9}
       \and
       Lindsey S. Wiser\inst{12}
       }
\institute{
    Université de la Côte d'Azur, Observatoire de la Côte d'Azur, CNRS, Laboratoire Lagrange, France
    \and
    Tsung-Dao Lee Institute \& School of Physics and Astronomy, Shanghai Jiao Tong University, Shanghai 201210, People's Republic of China
    \and
    Center for Space and Habitability, University of Bern, Gesellschaftsstrasse 6, CH-3012 Bern, Switzerland
    \and
    Department of Physics and Astronomy, Michigan State University, East Lansing, MI 48824, USA
    \and
    California Institute of Technology/IPAC, 1200 E. California Blvd, MC 100-22, Pasadena, CA 91125, USA
    \and
    Department of Astronomy, University of Wisconsin--Madison, Madison, WI 53703, USA
    \and
    AURA for the European Space Agency (ESA), Space Telescope Science Institute, 3700 San Martin Drive, Baltimore, MD 21218, USA
    \and
    Department of Astronomy and Astrophysics, University of California, Santa Cruz, CA 95064, USA
    \and
    School of Earth and Space Exploration, Arizona State University, Tempe, AZ, USA
    \and
    Division of Science, National Astronomical Observatory of Japan, 2-12-1 Osawa, Mitaka-shi 181-8588 Tokyo, Japan
    \and
    Steward Observatory, 933 North Cherry Avenue, Tucson, AZ 85721, USA
    \and
    Johns Hopkins Applied Physics Laboratory, Laurel, MD 20723, USA
    }   

   \date{}

  \abstract
   {JWST has shown a large diversity in warm Jupiter spectra, despite only small variations in the planetary parameters across the population. However, the main driver of this diversity remains unclear.}
   {We aim to identify the mechanisms responsible for the spectral difference of three warm Jupiter-size exoplanets observed by JWST: whereas WASP-80b appears mostly cloud-free, both WASP-107b and WASP-69b have spectra dominated by clouds.}
   {We model each planet using the same framework, ADAM (formerly SPARC/MITgcm), which solves for the interactions among cloud transport, radiative transfer, and atmospheric circulation in 3D. We investigate the role of three condensate species,  \nas, KCl, and \mg, and four particle sizes (0.1, 1, 5, and 10 \micron)}
   {Clouds settle deeper in the atmosphere of the higher-gravity planet WASP-80b than in WASP-107b, reproducing their spectral difference naturally. For WASP-107b, three clouds can reproduce the NIRCam observations: 5 \micron \nas, 1 \micron KCl, and 5 \micron \mg models. However, these cannot match the scattering slope observed at shorter wavelengths in NIRISS and the possible silicate feature in the MIRI bandpass, suggesting a multi-modal distribution of clouds. 
   Our model predicts that small silicate particles should be homogeneously distributed and thus cannot account for the difference between the two limb spectra in the MIRI bandpass. Finally, applying the same model to WASP-69b does not yield a partially cloudy dayside solution that fits the emission spectra, as proposed in a previous study.}
   {Coupling among 3D circulation, clouds, and radiative transfer can enhance the spectral diversity of warm Jupiter exoplanets, particularly through changes in cloudiness with gravity. The combination of multi-phase, wide-wavelength coverage and models that couple clouds, circulation, and radiative transfer is key to advancing our understanding of these new objects.}

   \keywords{planets and satellites: atmospheres - methods: numerical - infrared: planetary systems - planets and satellites: composition }
    \titlerunning{The tale of 3 planets}
   \maketitle

\section{Introduction}
\label{section: Introduction}

JWST now enables the cloud distribution of warm worlds ($T_{\rm eq}$ < 1000 K), revealing a striking diversity in their atmospheric spectra. These planets are essential laboratories for understanding atmospheric dynamics, chemistry, and cloud formation across diverse planetary environments. Clouds fundamentally shape exoplanetary atmospheres by controlling albedo, heat transport, and observed spectra, whereas circulation plays a key role in determining cloud distributions. However, the interplay among clouds, thermal structure, and three-dimensional atmospheric dynamics remains poorly understood in these transitional objects, and it remains unclear how it depends on planetary parameters or whether it can explain the observed spectral diversity.

Warm gas giants orbiting M-dwarf and K-dwarf stars are particularly valuable targets due to their high signal-to-noise ratios (SNR), which facilitate detailed characterization of atmospheric composition, thermal structure, and aerosol properties. The high-quality spectra achievable for these systems allow for unprecedented constraints on cloud particle sizes, compositions, and spatial distributions, properties that remain poorly understood for most exoplanets.

The diversity of clouds in observations can be highlighted by the observations of 3 warm giants: WASP-107b \citep{welbanks2024}, WASP-80b \citep{wiser2025, mehta2026, triantafillides2026}, and WASP-69b \citep{schlawin2024}.
Indeed, although they share similar radius ($\sim$1 $R_{\rm Jup}$) and equilibrium temperatures (725-935 K), their observations reveal striking differences in their atmospheric spectrum. 
This suggests that subtle variations in planetary mass, host-star properties, and orbital parameters can lead to markedly different atmospheric states. These variations may arise from the complex interplay among cloud microphysics, radiative transfer, and atmospheric dynamics, whose nonlinear interactions can amplify the atmosphere's response to small changes in planetary parameters.
All three planets have been observed as part of the MANATEE program (JWST-GTO-1185, PI: T. Greene; \citealt{schlawin2018}), providing a homogeneous dataset ideal for comparative studies. The detailed planetary and stellar parameters are summarized in Table \ref{tab:planet_params}.

\begin{table*}
\caption{Physical and orbital parameters of the studied planets.}
\label{tab:planet_params}
\centering
\begin{tabular}{lccc}
\hline\hline
Parameter & WASP-107b & WASP-80b & WASP-69b \\
\hline
Planetary radius ($R_p$) [$R_{\mathrm{J}}$] & 0.948  & 0.978  & 1.057 \\
Planetary mass ($M_p$) [$M_{\mathrm{J}}$]   & 0.12   & 0.538  & 0.26  \\
Gravity ($g$) [m s$^{-2}$]                 & 2.82   & 13.96  & 5.32  \\
Equilibrium temperature ($T_{\mathrm{eq}}$) [K] & 735 & 825 & 962 \\
Stellar effective temperature ($T_{\mathrm{eff}}$) [K] & 4430 & 4143 & 4715 \\
Orbital period ($P$) [days]                 & 5.721  & 3.068  & 3.868 \\
Semi-major axis ($a$) [AU]                  & 0.0344 & 0.0558 & 0.0453 \\
Metallicity ($[M/H]$)                       & 1.09 & 0.55 & 1.0 \\
$C/O$                                       & 0.33 & 0.48 & 0.55 \\
Internal effective temperature $T_{\rm int}$ (K) $^*$
& [500, 500, 500] 
& [100, 100, 381$^{\triangle}$] 
& [100, 100, 100] \\
\hline
\end{tabular}
\tablefoot{Planetary and stellar parameters were taken from \cite{welbanks2024,wiser2025,schlawin2024}. For internal effective temperature (T$\rm _{int}$), see section \ref{subsection: tint}. $^*$ The $T_{\rm int}$ corresponds to [\nas, KCl, \mg] respectively. $^{\triangle}$ See \cite{mehta2026} for detailed explanation. }
\end{table*}

\begin{figure*}[!hbt]
\centering
\includegraphics[width=\textwidth]{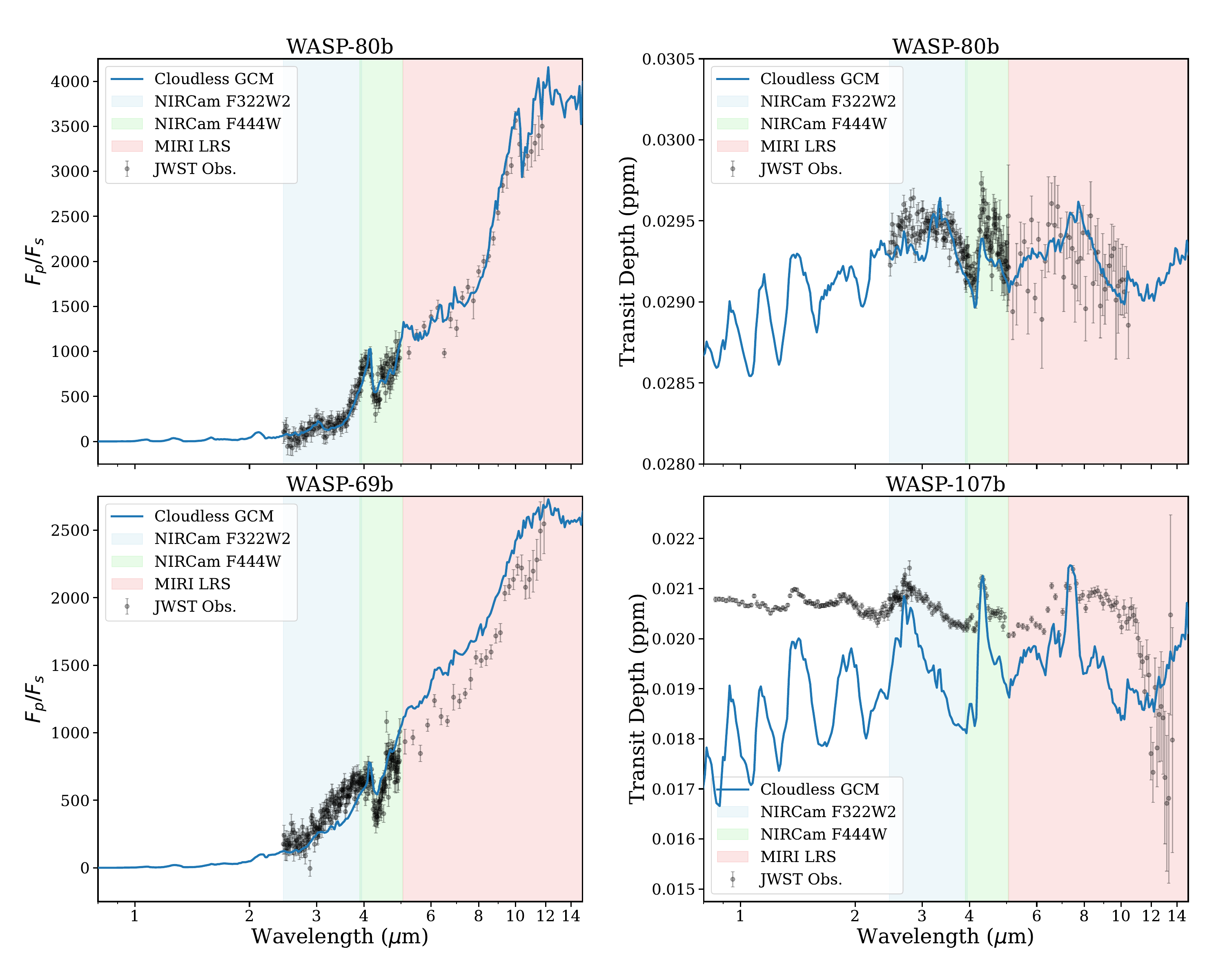}
\caption{Cloudless GCM spectra compared with JWST observations. \textit{Top left:} Emission spectrum of WASP-80b. \textit{Top right:} Transmission spectrum of WASP-80b. \textit{Bottom left:} Emission spectrum of WASP-69b. \textit{Bottom right:} Transmission spectrum of WASP-107b. In each panel, black points show JWST observations.}
\label{fig:transmission_emission}
\end{figure*}

Figure \ref{fig:transmission_emission} presents JWST transmission and emission spectra for three warm gas giants alongside cloudless GCM predictions, providing a uniform framework for comparative atmospheric analysis. The use of identical instrumentation and consistent data reduction across all targets minimizes systematic differences, making this dataset particularly well-suited to isolating genuine atmospheric diversity.

All three planets exhibit common molecular absorbers, including H$_2$O bands near 2.8 \micron and 6.7-8.0 \micron, and CO$_2$ at 4.3 \micron, with excess absorption on the red wing consistent with CO near 4.7 \micron. However, the relative prominence of these features varies, underscoring the influence of atmospheric structure and composition.

Methane (CH$_4$) provides a useful point of comparison across this sample. At the temperature of WASP-69b ($\sim$963 K), CH$_4$ is expected to be abundant under chemical equilibrium at $\sim$0.1 bar for near-solar composition \citep{schlawin2024}. However, strong CH$_4$ features near 3.3 \micron and 7.7 \micron are not observed. Similarly, WASP-107b exhibits a depleted CH$_4$ abundance due to high $T_{\mathrm{int}}$ and quenching. In contrast, WASP-80b exhibits prominent CH$_4$ absorption. This diversity highlights the potential roles of disequilibrium chemistry, vertical mixing, and cloud opacity in shaping methane observability.
Sulfur-bearing species provide an additional comparative diagnostic. Evidence for SO$_2$ has been reported in WASP-107b, consistent with photochemical processing in a highly irradiated, low-gravity atmosphere. WASP-80b, on the other hand, shows signatures attributed to CS$_2$ \citep{mehta2026,veillet2025, triantafillides2026}, suggesting a distinct chemical pathway. For WASP-69b, sulfur species remain less clearly constrained, potentially reflecting weaker features or obscuration by additional opacity sources. Together, these differences indicate a diversity of sulfur-chemistry regimes across planets with otherwise comparable bulk properties. 

Despite their broadly similar radii and equilibrium temperatures, the three planets exhibit notably different spectra. WASP-80b is broadly consistent with cloudless model predictions, suggesting relatively unobscured molecular features \citep{mehta2026}. In contrast, WASP-107b displays significantly muted spectral structure compared to the cloudless case, indicative of substantial opacity from high-altitude clouds or hazes \citep{welbanks2024, sing2024, dyrek2024}. WASP-69b exhibits intermediate behavior \citep{schlawin2024}: while some features, such as CO$_2$ at 4.3 \micron, are reproduced by cloudless models, the overall spectral shape suggests that additional mechanisms, such as clouds, are required.

This comparative sample, therefore, provides an opportunity to investigate the role of clouds as a primary driver of atmospheric diversity in warm gas giants. Variations in cloud properties, such as composition, particle size, and vertical distribution, can significantly shape the observed spectra. Constraining these effects is essential for interpreting the growing population of exoplanet spectra now accessible with JWST.

In this work, we build upon the framework of \citet{mehta2026} to conduct a comprehensive comparative study of WASP-107b, WASP-80b, and WASP-69b using three-dimensional General Circulation Models with radiatively active clouds. We aim to explain the diverse atmospheric properties observed in these three warm gas giants through a unified modeling framework that incorporates cloud formation, transport, and radiative feedback processes. The clouds are modeled as dynamical tracers distributed according to atmospheric circulation patterns and vertical settling velocities derived from GCMs. These cloud tracers provide radiative feedback to the atmospheric temperature structure, allowing us to investigate how clouds actively shape the thermal profiles and spectral appearance of each planet. By comparing our GCM predictions with existing JWST observations across all three targets, we seek to constrain the cloud properties, including composition, particle sizes, and spatial distributions, that best reproduce the observed emission and transmission spectra. Furthermore, we predict future observations of these planets that can break the degeneracy among different cloud scenarios and provide additional constraints on atmospheric dynamics and chemistry.

The paper is structured as follows.
Section \ref{section: methods} describes the atmospheric composition of the planets adopted in the GCM, the three-dimensional GCM framework, the cloud tracer scheme, and the post-processing methodology.
Section \ref{section: cloudless} presents the cloudless modeling results for each planet and compares synthetic spectra against observations.
Section \ref{section: cloudy} introduces cloudy models, beginning with a general overview, followed by a detailed analysis of WASP-107b, a comparative study of WASP-107b and WASP-80b, a limb-asymmetry analysis, and finally the WASP-69b models.
Section \ref{section:conclusion} summarizes our conclusions.

\section{Methods}
\label{section: methods}

In this section, we describe the three-dimensional GCM framework, the implementation of the cloud tracer scheme within the GCM, and the post-processing methodology applied to the GCM outputs.

\subsection{Atmospheric composition}
\label{section: Atmospheric composition}

Warm Jupiters are less likely to follow thermochemical equilibrium compared to hot Jupiters, primarily due to their lower atmospheric temperatures. At these cooler temperatures, chemical reaction rates are significantly slower, allowing disequilibrium processes such as vertical mixing and photochemistry to dominate the chemical composition \citep{mukherjee2025}. In contrast, the higher temperatures on hot Jupiters enable faster reaction rates, allowing the atmosphere to maintain equilibrium despite dynamical transport. As a result, warm Jupiter atmospheres often show deviations from equilibrium predictions \citep{drummond_implications_2020, zamyatina2022, zamyatina_quenching-driven_2024, lee2023a}.

To reduce the parameter space explored in our GCM simulations, we adopt chemical abundances based on existing atmospheric retrieval studies for each planet. The atmospheric compositions for the three planets are adopted from WASP-107b \citep{welbanks2024}, WASP-69b \citep{schlawin2024}, and WASP-80b \citep{wiser2025,mehta2026}, with the corresponding metallicities listed in Table \ref{tab:planet_params}.
Further planetary and stellar parameters are also mentioned in the Table \ref{tab:planet_params}. This method successfully reproduced the transmission and emission observations for WASP-80b \citep{mehta2026}.
The retrieved chemical abundances vary with pressure, reflecting the combined effects of thermochemistry, photochemistry, and vertical quenching. We assume that horizontal mixing is efficient enough to produce horizontally homogeneous chemical abundances at each pressure level. Therefore, we apply one-dimensional, vertically varying abundance profiles to each GCM column for each planet. 
The typically fast horizontal transport timescales support this assumption compared to vertical mixing in tidally locked exoplanet atmospheres \citep{showman2009,zhang2017}. However, we note that the morning and evening limbs may exhibit different chemical compositions \citep{murphy2025}, implying that our models are not expected to fully reproduce the observed limb transmission spectrum of WASP-107b (see Section \ref{section:la}).
Clouds, however, are implemented as tracers in both the gas phase and condensed phase (see Section \ref{subsec:clouds}), allowing their spatial distribution to be fully inhomogeneous and determined self-consistently by the atmospheric dynamics.

We follow the opacity treatment described in \citet{mehta2026}. 
To accommodate disequilibrium-chemistry abundances, we use a custom module that computes opacity tables by mixing correlated-k tables from \citet{freedman2014,gharib-nezhad2024} according to the retrieved chemical abundances for each planet.
We employ the \texttt{exo\_k} Python library \citep{leconte2021} to bin the opacity data into 11 spectral bands following \citet{kataria2016}, and combine the k-coefficients using the random overlap method of \citet{lacis1991} (see Appendix \ref{appendix:opa_mix} for a discussion of the bin-then-mix versus mix-then-bin approaches).
The resulting opacity tables are fixed throughout each GCM simulation and are planet-specific, reflecting the different atmospheric compositions of WASP-107b \citep{welbanks2024}, WASP-80b \citep{wiser2025, mehta2026}, and WASP-69b \citep{schlawin2024}.

\subsection{Internal Temperature}
\label{subsection: tint}

Following \citet{mehta2026}, we account for both incoming stellar irradiation and intrinsic planetary heat flux in our models. The latter is parameterized through an internal effective temperature $T_{\rm int}$, which has been shown to vary significantly among hot and warm Jupiters, potentially due to processes such as ohmic dissipation \citep{batygin2010, guillot2023, thorngren2019, fortney2021}.

For our cloudless atmospheric models, we explore two values of $T_{\rm int}$ for each planet: a low $T_{\rm int}$ = 100 K baseline representing standard cooling predictions \citep{guillot1996}, and a high $T_{\rm int}$ value motivated by atmospheric retrieval studies. For WASP-80b, we adopt $T_{\rm int}$ = 381 K from \citet{wiser2025}; for WASP-107b and WASP-69b, we use the value $T_{\rm int}$ = 500 K \citep{welbanks2024}. For models including radiatively active clouds, we fix $T_{\rm int}$ to the values derived from the respective one-dimensional retrieval studies for each planet as shown in the Table \ref{tab:planet_params}. 
The cloudless cases with low $T_\mathrm{int}$ for WASP-107b and high $T_\mathrm{int}$ for WASP-69b are not discussed as they have very similar 3D thermal and dynamical structure; however, for WASP-80b, changes are observed in the dynamical structure \citep{mehta2026}.

\subsection{General Circulation Model}
\label{subsec:gcm}

We use ADAM (ADvanced Atmospheric MITgcm) with the SPARC radiative transfer module, active tracer clouds, and custom abundance modules to simulate the atmospheres of WASP-107b, WASP-80b, and WASP-69b. ADAM is an umbrella designation for a suite of exoplanet modeling frameworks built upon the MITgcm \citep{adcroft2004}, coupling atmospheric dynamics with plane-parallel radiative transfer \citep{marley1999}. The details of ADAM, including its application to various exoplanet classes, are described in \citet{mehta2026} and references therein.

We follow the numerical setup of \citet{mehta2026}, using a cubed-sphere grid with C32 resolution (128$\times$64 in longitude$\times$latitude) and 53 vertical pressure levels spanning from 200 bar to 2 $\mu$bar, providing approximately three levels per atmospheric scale height. Temperature profiles are initialized using the analytical model of \citet{parmentier2015}, and we apply a fourth-order horizontal Shapiro filter for numerical stability \citep{showman2009, koll2018,parmentier2021}. No explicit Rayleigh drag is included. 
We also include a convective scheme in all models because the deep layers reach the unstable convective region. The effects of rapid convective mixing were parametrized using a simple convective adjustment scheme as in the NCAR Community Atmosphere Model (\cite{collins2004}, see their Section 4.6).
The planet-specific parameters (gravity, radius, orbital period, semi-major axis, and metallicity) for each target are listed in Table \ref{tab:planet_params}.

\subsection{Cloud Tracer Scheme}
\label{subsec:clouds}

Clouds are implemented as radiatively active tracers that evolve according to local temperature, pressure, and atmospheric circulation, following the methodology of \citet{tan2021,tan2021a,komacek2022a} and following the same setup as in \cite{mehta2026}. We include Na$_2$S, KCl, and MgSiO$_3$ clouds for all three planets, as these species are expected to condense at the relevant atmospheric temperatures based on their condensation curves \citep{visscher2006,morley2012,visscher2010}. 
Each cloud species is initialized with the solar element ratios of its dominant condensate species \citep{lodders2003}, scaled to the atmospheric metallicity of each planet (details in Table \ref{tab:cloud-properties}).

The cloud tracers are governed by coupled equations for condensible vapor ($q_v$) and cloud condensate ($q_c$) that account for condensation, evaporation, vertical settling, and deep vapor replenishment. Cloud particles follow a log-normal size distribution with width $\sigma = 1.65$, and we explore mean particle sizes of $r_0$ = 0.1, 1, 5, and 10 $\mu$m for each cloud species. Settling velocities are calculated following \citet{parmentier2013a}. The radiative feedback of clouds is computed from Mie scattering theory using optical constants, as detailed in \citet{mehta2026}. The cloud microphysical parameters (deep mixing ratios, source pressures, and relaxation timescales) are identical to those used in \citet{mehta2026} and are provided in Table \ref{tab:cloud-properties}.

\begin{table*}[!hbt]
\centering
\setlength{\tabcolsep}{12.5pt}
\caption{Input parameters for cloud properties in the GCM simulations.}
\begin{tabular}{llll}
\hline
\hline
\multicolumn{4}{c}{Cloud properties} \\ \hline
& Na$_2$S & KCl & MgSiO$_3$ \\ \hline

Condensate vapor deep mixing ratio ($q_{\rm {deep}}$) 
& $5.9 \times 10^{-5} \times 10^{[M/H]}$ 
& $7.13 \times 10^{-6} \times 10^{[M/H]}$ 
& $0.0027 \times 10^{[M/H]}$ \\

Condensate vapor source pressure ($p_{\rm {deep}}$) 
& $10^7$ 
& $10^7$ 
& $3.3 \times 10^7$ \\

Condensate vapor relaxation timescale ($\tau_{\rm {c}}$) 
& 100 s 
& 100 s 
& 200 s \\

Condensate vapor deep relaxation timescale ($\tau_{\rm {deep}}$) 
& 1000s 
& 1000s 
& 100 s \\

Condensate density ($\rho_c$) \citep{roman2021}
& 1860 kg/m$^3$ 
& 1980 kg/m$^3$ 
& 3190 kg/m$^3$ \\

Mean particle size ($r_0$) 
& [0.1, 1, 5, 10] $\mu$m 
& [0.1, 1, 5, 10] $\mu$m 
& [0.1, 1, 5, 10] $\mu$m \\

Log-normal distribution width ($\sigma$) 
& 1.65 
& 1.65 
& 1.65 \\ \hline

\end{tabular}

\tablefoot{
Metallicity is represented by [M/H], with values from Table \ref{tab:planet_params}.  
Optical constants are taken from: Na$_2$S - \citet{Montaner1979, Khachai2009},  
KCl - \citet{Querry1987OpticalCO},  
MgSiO$_3$ - \citet{Scott1996}.
}

\label{tab:cloud-properties}
\end{table*}

\subsection{Post-processing}
\label{subsec:postprocessing}

We post-process our GCM outputs using \texttt{gcm\_toolkit} \citep{carone2020,schneider2022}, an open-source Python package that regrids the cubed-sphere output to a regular latitude-longitude grid of (45, 72) while maintaining the 53 vertical pressure levels. The regridded atmospheric structure is then used as input to \texttt{gCMCRT} \citep{lee2022}, a three-dimensional GPU-accelerated Monte Carlo radiative transfer code that operates on spherical geometry. \texttt{gCMCRT} has been successfully applied to ADAM outputs in previous studies \citep{komacek2022a,wardenier2021,mehta2026} and produces synthetic transmission and emission spectra by tracing photon packets through the 3D atmospheric structure. Limb-resolved spectra are computed using the 3D radiative transfer code \texttt{gCMCRT}, which tracks photons passing through each limb separately \citep{wardenier2021}.

Unlike plane-parallel radiative transfer codes that produce flux per unit surface area, \texttt{gCMCRT} directly computes the total outgoing flux, accounting for the three-dimensional geometry and volume of emitting regions. 
The wavelength-dependent effective radius is already implicitly encoded in the Monte Carlo volume-integrated treatment of the atmosphere. By integrating directly over the full emitting volume, gCMCRT naturally includes contributions from all atmospheric layers, with hotter and cooler regions weighted according to their physical emission properties. This avoids biases that can arise when multiple one-dimensional temperature–pressure profiles are combined into a single spectrum \citep{feng2016,taylor2021}.

The treatment of planetary radius differs between our three targets. For WASP-80b, we follow the benchmarking procedure described in \citet{mehta2026}. For WASP-107b, we calculate a transit spectrum with \texttt{gCMCRT} with our initial 200 bar radius (0.948 R$_J$). We then determine the radius difference between the observed and modeled radius and adjust our 200 bar radius accordingly, so that the calculated transit spectra match the observations. Whereas for WASP-69b, since there is no transmission spectrum, the planetary radius is treated as a free parameter and adjusted to match the observed spectra for each model. 

For cloudy models, cloud particle sizes follow a log-normal distribution with width $\sigma = 1.65$ and mean particle radius $r_0$ as specified in Table \ref{tab:cloud-properties}. The particle size distribution affects both the radiative properties computed by \texttt{gCMCRT} and the vertical settling velocities in the GCM through the terminal velocity term in the cloud condensate equation (see Section \ref{subsec:clouds} and \citealt{mehta2026} for details).

\section{Cloudless GCMs}
\label{section: cloudless}

\begin{figure*}[!hbt]
\centering
\includegraphics[width=0.9\textwidth]{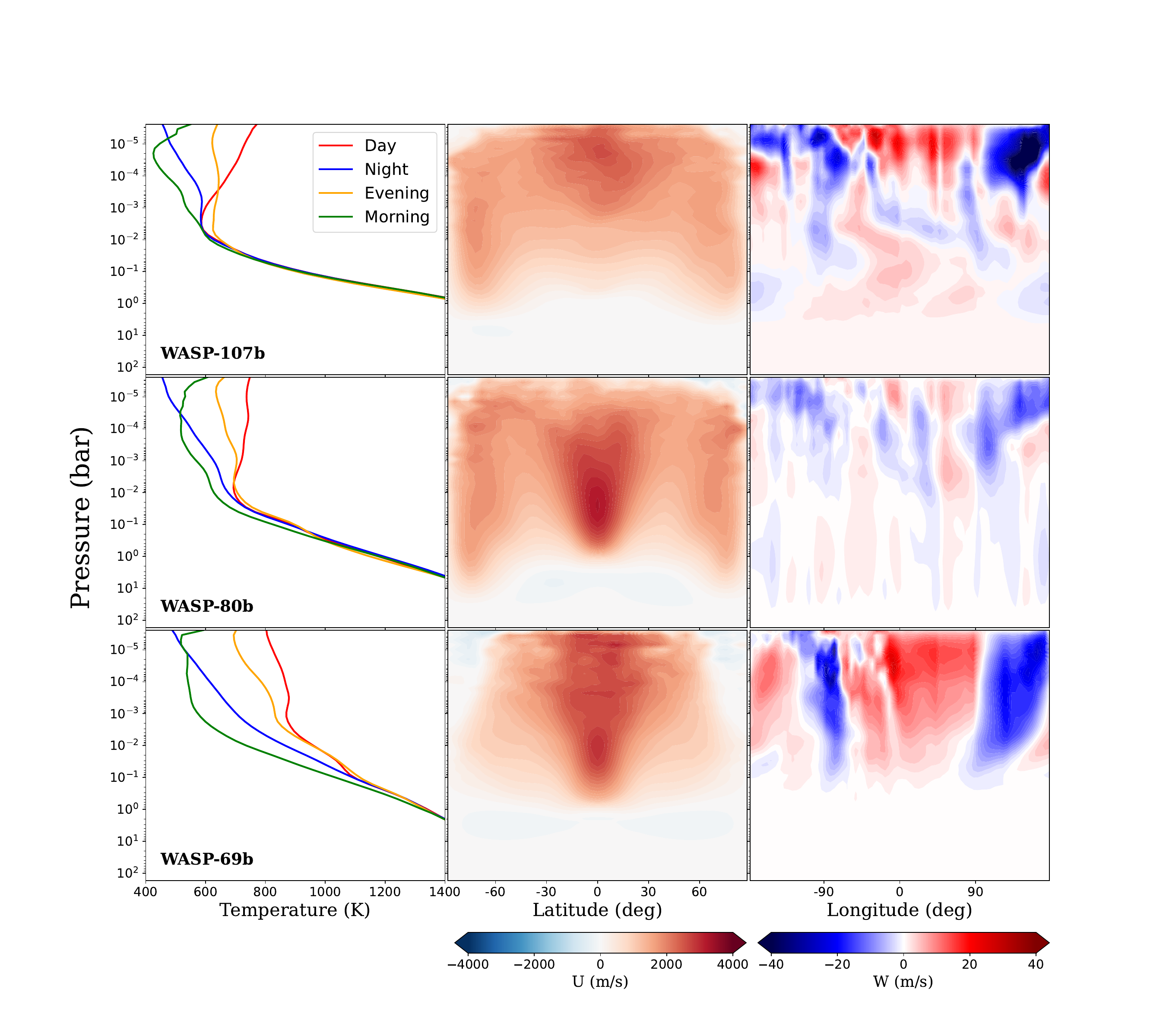}
\caption{
Comparison of cloudless general circulation models for WASP-107b, WASP-80b, and WASP-69b.
The left column shows temperature-pressure profiles for different regions of the planet (dayside, nightside, morning, and evening terminators).
The middle column shows the zonal-mean zonal wind speed $U$, corresponding to the latitudinal distribution of the zonal wind component averaged over longitude.
The right column shows the zonal-mean vertical wind speed $W$, corresponding to the longitudinal distribution of the vertical wind component averaged over latitude.
Positive values of $W$ indicate upwelling and negative values indicate downwelling, highlighting the day-night circulation pattern in the atmospheres of the three planets.
}
\label{fig:cldless}
\end{figure*}

We use ADAM to model the cloudless atmospheres of WASP-107b, WASP-80b, and WASP-69b to compare their thermal structures, atmospheric dynamics, and spectral behavior.

WASP-107b is the planet with the lowest temperature and gravity among the three planets. Observations indicate that it has a high $T_{\mathrm{int}}$ \citep{welbanks2024, sing2024}. Fig. \ref{fig:cldless} (top row) shows a slight temperature inversion in the low-pressure regions. The temperature difference between the eastern and western limbs is approximately 200 K, indicating the influence of atmospheric dynamics and opacity on limb asymmetry. The circulation is dominated by a near-prograde zonal wind extending across most latitudes, with weak vertical winds. However, the cloudless GCM produces very large features in the transmission spectrum compared to the observations, clearly indicating the necessity of clouds in its atmosphere (Fig. \ref{fig:transmission_emission}).

As seen in Fig. \ref{fig:cldless} (middle row), WASP-80b exhibits an efficient day–night heat redistribution in the cloudless case. The temperature contrast at low pressures decreases rapidly with depth, and the atmosphere becomes nearly horizontally homogeneous at higher pressures. The circulation is characterized by a dominant equatorial super-rotating jet with prograde winds extending across latitudes. The vertical winds are not strong and lack structure.
As shown in Fig. \ref{fig:transmission_emission}, the cloudless GCM reproduces the emission spectrum well, although discrepancies remain around specific wavelength regions. In transmission, the model provides a reasonable match but slightly overestimates certain spectral features \citep{mehta2026}. 

WASP-69b is the hottest of the three planets. Similar to the other two planets, it exhibits a dominant near-prograde equatorial jet. 
However, unlike WASP-107b and WASP-80b, retrograde winds develop in the polar regions, accompanied by strong upwelling on the dayside and downwelling on the nightside (Fig. \ref{fig:cldless}).
One possible explanation for this distinct polar circulation lies in the effect of reduced atmospheric static stability in hotter atmospheres. However, further dedicated dynamical studies are required to confirm this interpretation.
Whereas WASP-107b and WASP-80b adopt chemical abundances informed by one-dimensional atmospheric studies (\citealt{welbanks2024, wiser2025} resp.), WASP-69b assumes equilibrium chemistry throughout \citep{schlawin2024}. 
These differing chemical prescriptions alter the local opacity and radiative timescales, thereby modifying the atmospheric thermal structure. 
In particular, the assumption of equilibrium chemistry in WASP-69b may lead to a steeper effective lapse rate, thereby reducing atmospheric static stability. This reduction is quantified through the Brunt--V\"{a}is\"{a}l\"{a} frequency, $N$, defined as
\begin{equation}
    N^2 = \frac{g}{\theta}\frac{d\theta}{dz},
    \label{eq:BV}
\end{equation}
where $g$ is the gravitational acceleration, $\theta$ is the potential temperature, and $z$ is the altitude, measures the degree of stable stratification of the atmosphere. 
In the limit of an isothermal atmosphere, $N\approx g/\sqrt{Tc_p}$, whereas it is zero when the thermal profile is convective. In between, $N$ decreases as the thermal gradient becomes more negative.
The Rossby deformation radius, defined as
\begin{equation}
    L_D = \frac{NH}{f},
    \label{eq:Ld}
\end{equation}
where $H$ is the atmospheric pressure scale height and $f = 2\Omega\sin\phi$ is the Coriolis parameter at latitude $\phi$, with $\Omega$ the planetary rotation rate, sets the characteristic horizontal scale over which rotational effects become dynamically important. 
When $L_D$ is large compared to the planetary radius, a single equatorial jet dominates the global circulation, as seen in the cooler WASP-107b and WASP-80b. 
In WASP-69b, however, the assumption of equilibrium chemistry leads to a larger vertical temperature gradient, thereby lowering both $N$ and $L_D$ and explaining the presence of a counter-rotating polar jet in our WASP-69b model. 
In the cloudless case, the GCM reproduces the 4.3 $\mu$m absorption feature. The model underestimates the flux at shorter wavelengths and overestimates it at longer wavelengths, as shown in Fig. \ref{fig:transmission_emission}.

All three planets display predominantly prograde equatorial flows in the cloudless simulations. However, they differ in their thermal structures and dynamics. These differences illustrate how variations in planetary properties lead to distinct atmospheric structures and observable signatures in the cloudless regime (Fig. \ref{fig:transmission_emission}). 

\section{Cloudy GCMs}
\label{section: cloudy}
\subsection{Clouds: General Overview of the Three Planets}

We modeled the atmospheres of WASP-107b, WASP-80b, and WASP-69b, including different cloud species and particle sizes, resulting in a grid of cloudy GCM simulations. A common trend across all three planets is that smaller particles are present at lower pressures. As the particle size increases, gravitational settling becomes more efficient, and the cloud decks shift toward deeper pressure levels for all species (\citealt{mehta2026} and Fig. \ref{fig:cldna2s}, \ref{fig:cldkcl}, \ref{fig:cldsil}).
The lower cloud boundary is set by the pressure at which the condensation curve intersects the local temperature-pressure (T-P) profile. The particularly deep 1 \micron \mg case in WASP-107b (Fig. \ref{fig:cldsil}) results from a change in the thermal structure caused by the reflective clouds, which shifts this intersection to higher pressures relative to the other cases. Similarly, in WASP-69b, all \mg cases exhibit deep cloud decks.
Clouds are implemented as fully coupled tracers in the GCM, and their spatial distribution is therefore controlled by the joint effect of the circulation and the thermal structure. The equatorial super-rotating jet efficiently redistributes condensates, leading to equatorial depletion through zonal advection and associated vertical motions. This depletion is present in all three planets. However, as seen in Fig. \ref{fig:cldsil}, small \mg clouds (0.1 and 1 \micron) in WASP-107b remain sufficiently mixed to resist equatorial clearing partially. A similar, but weaker, behavior is observed in WASP-69b.
Dayside evaporation is observed in most cases due to the stellar heating. An exception is found for the \mg cases on all three planets, where upper atmospheric temperatures do not exceed the condensation threshold, and significant evaporation does not occur (shown in Fig. \ref{fig:cldsil}).

The thermal impact of clouds depends on their abundance, vertical distribution, and optical properties. When clouds are confined to deep atmospheric layers, they do not significantly modify the T-P profile. In these regions, long radiative timescales damp temperature perturbations, and the stellar flux is primarily absorbed at higher altitudes. By contrast, small cloud particles are efficiently mixed at all pressures and can therefore significantly impact the thermal structure throughout the atmosphere.

\nas clouds show the strongest radiative feedback owing to their high optical depth and shortwave absorption. Small \nas particles (0.1 \micron) are very efficient at absorbing stellar radiation, producing pronounced thermal inversions above the cloud layer and shielding and cooling the atmosphere below (anti-greenhouse effect; \citealt{guillot2010}). Larger \nas particles mainly contribute to local heating depending on their abundance and vertical location. This behavior is evident in WASP-69b (Fig. \ref{fig:tna2s}), where \nas clouds are less abundant, and temperature changes are moderate, whereas WASP-80b and WASP-107b exhibit stronger thermal modifications.

KCl clouds are generally optically thinner and have weaker shortwave absorption than \nas, so their impact on the temperature structure (Fig. \ref{fig:tkcl}) is limited unless their abundance is high. A substantial KCl abundance is found in WASP-107b (Fig. \ref{fig:cldkcl}), leading to moderate heating, while the other two planets show negligible temperature changes.

\mg clouds are optically thick and, depending on particle size, display both reflective and absorptive properties. Small \mg particles (0.1 and 1 \micron) are present at low pressures and induce thermal inversions (Fig. \ref{fig:tsil}) similar to those produced by \nas. In contrast, larger particles reside deeper in the atmosphere and primarily contribute to local heating without significantly altering the overall T-P structure.

Clouds influence atmospheric circulation through radiative-dynamical coupling. By modifying the heating rates, they alter horizontal temperature gradients and thus the pressure gradients that drive the winds. The dynamical response, therefore, arises from cloud-induced radiative feedback.
A correlation between strong vertical winds and the presence of a well-defined equatorial jet is observed in these and other cloud scenarios (see also \citealt{mehta2026}). Models with enhanced vertical winds also possess a pronounced equatorial jet accompanied by retrograde winds at higher latitudes.  
Figure \ref{fig:pkzz} presents the vertical eddy diffusion coefficient, K$_{zz}$, for all GCM simulations. 
The K$_{zz}$ profiles corresponding to the different cloud cases are computed following $K_{zz}(p)\sim H \cdot w_{\rm rms}(p)$, where $H$ is the atmospheric scale height and $w_{\rm rms}$ is the root-mean-square vertical wind speed as a function of pressure \citep{lewis2010, moses2011}. 
Although this method generally overestimates the mixing efficiency by one to two orders of magnitude \citep{parmentier2013a}, it remains useful for comparing the relative mixing strengths among the models. We find that WASP-107b and WASP-69b have higher K$_{zz}$ than WASP-80b, and that changes in K$ _ {zz} $ with clouds and their particle sizes are observed in the models.
For WASP-107b, all \nas and \mg cases produce a well-defined equatorial jet together with strong vertical winds (Fig \ref{fig:uwna2s}, \ref{fig:uwsil}), whereas KCl clouds induce a weaker dynamical response due to their limited radiative impact (Fig. \ref{fig:uwkcl}). In WASP-69b, most models display strong vertical winds and a clear equatorial jet, except for the 0.1 \micron \nas case, which exhibits predominantly prograde winds across all latitudes (Fig. \ref{fig:uwna2s}). This behavior is consistent with the modified temperature gradients associated with the thermal inversion in that case.

\subsection{WASP-107b: Cloud models}

\begin{figure*}[!hbt]
\centering
\includegraphics[width=\textwidth]{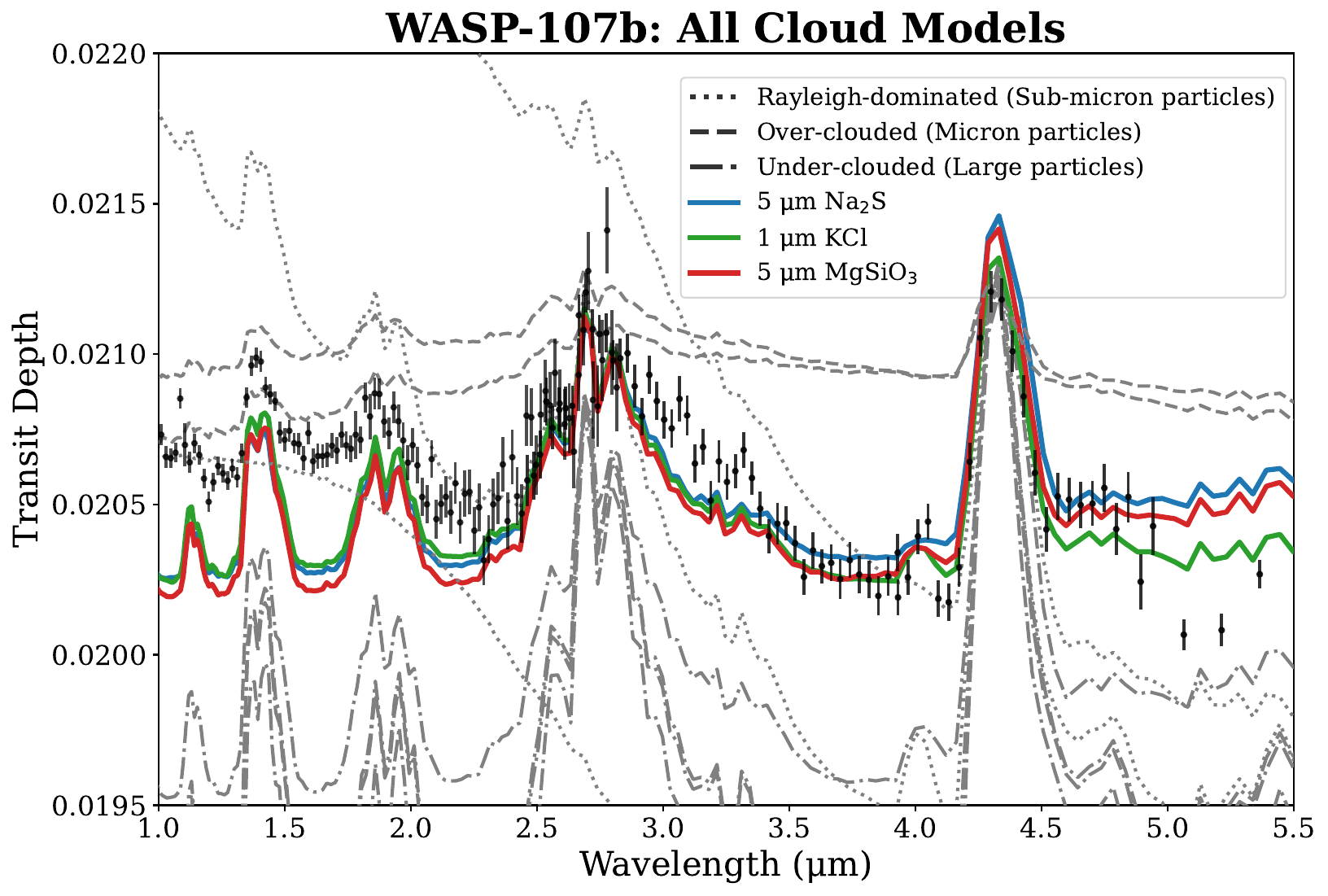}
\caption{Transmission spectra of all cloud models for WASP-107b compared with the JWST NIRCam observations (black points).
    Dotted lines show small-particle models dominated by Rayleigh scattering (0.1 \micron \nas, KCl, \mg); dashed lines are over-clouded 
    models with suppressed molecular features (1 \micron \nas and \mg); dot-dashed lines are under-clouded models with large features (10 \micron \nas; 5, 10 \micron KCl; 10 \micron \mg). Solid lines indicate the three best-fit models: 5 $\mu$m Na$_2$S (blue), 1 $\mu$m KCl (green), 
    and 5 $\mu$m MgSiO$_3$ (red).}
\label{fig:reject107}
\end{figure*}

\begin{figure*}[!hbt]
\centering
\includegraphics[width=\textwidth, trim=5 5 5 5, clip]{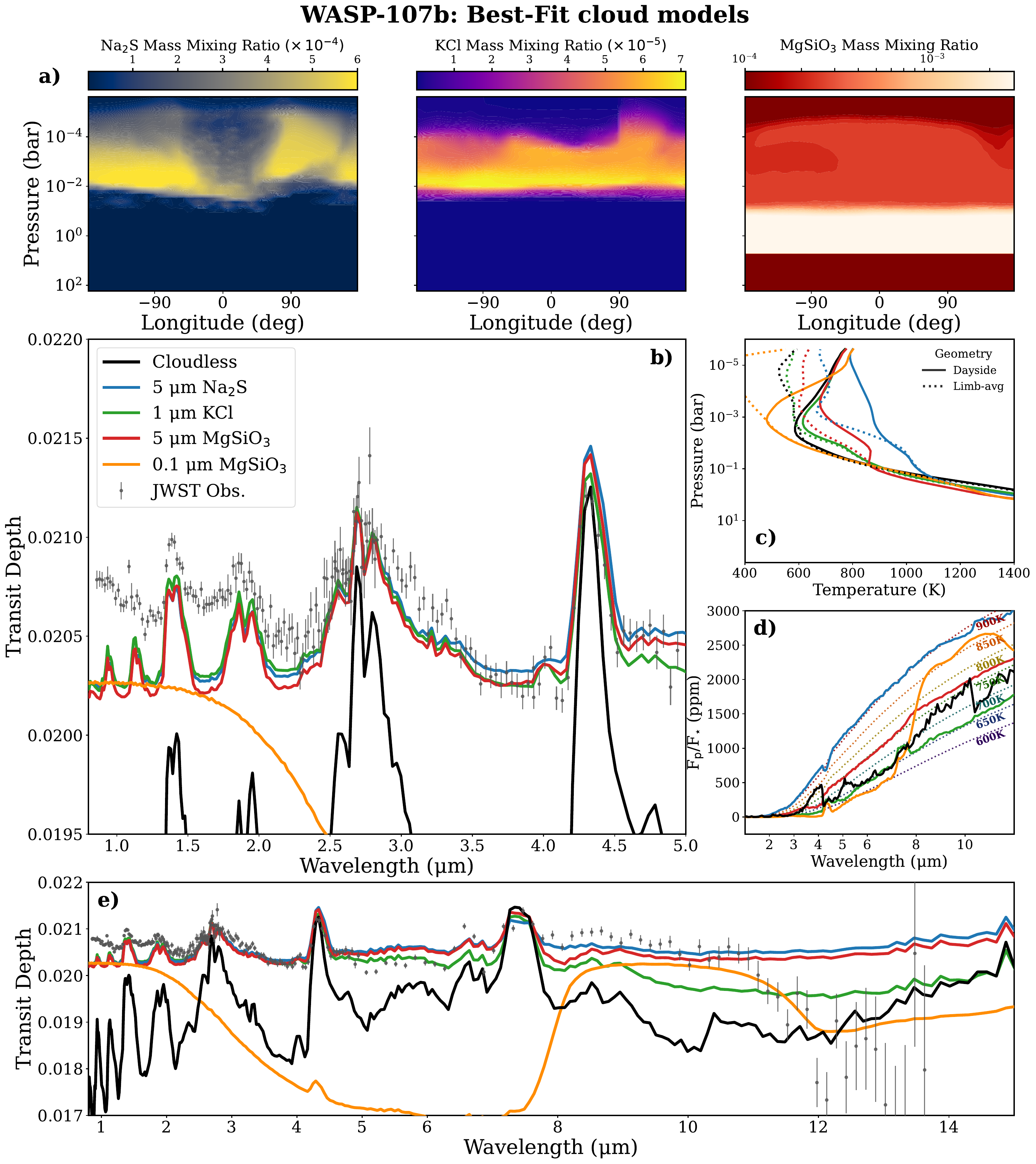}
\caption{Panel \textbf{a} shows the longitudinal distribution of Na$_2$S (left), KCl (middle), and MgSiO$_3$ (right) clouds (showing dayside and nightside), averaged over latitudes for WASP-107b models. Panel \textbf{b} shows the transmission spectra of the 3 best-fit models along with the JWST observation for NIRCam. Panel \textbf{c} shows corresponding limb-averaged (dotted) and dayside averaged (solid) TP profiles with emission spectra in panel \textbf{d}. Panel \textbf{e} shows NIRCam + MIRI observations with the models.}
\label{fig:spec_best3}
\end{figure*}

We computed a grid of cloud models for WASP-107b, varying both the cloud species and particle size. As illustrated in Fig. \ref{fig:reject107}, the resulting transmission spectra exhibit distinct behaviors depending on the cloud properties. Models with 0.1 \micron \nas, KCl, and \mg particles produce a pronounced Rayleigh scattering slope at short wavelengths alongside an elevated continuum at longer wavelengths. 
Notably, 0.1 \micron \mg clouds reproduce a spectral feature broadly consistent with the MIRI observations, suggesting that \mg can serve as a proxy for silicate clouds composed of mixtures such as SiO and SiO$_2$ \citep{dyrek2024}, and that a proper treatment of such mixed silicate compositions would likely reproduce the MIRI feature more faithfully.
At 1 \micron, \nas and \mg clouds extend throughout a large portion of the atmosphere, suppressing the molecular feature amplitudes smaller than the observed values. In contrast, KCl clouds of the same size provide a satisfactory match to the data. At 5 \micron, \nas and \mg clouds likewise reproduce the observed spectrum, whereas KCl particles of this size produce an excessively deep continuum inconsistent with the observations. Finally, particles with radii of 10 \micron or larger settle to deeper atmospheric layers, partially muting the spectral features, in contrast to WASP-80b, where large particles have a negligible effect on the transmission spectrum \citep{mehta2026}, yet still fail to reproduce the observed amplitudes.

We identify cloud models that partially reproduce the observed transmission spectrum.
As shown in Fig. \ref{fig:spec_best3}, within the explored parameter space, only models that include 5 \micron \nas, 1 \micron KCl, and 5 \micron \mg reproduce the NIRCam observations, capturing the correct cloud-deck opacity and spectral slope across the near-infrared.
Independently, 0.1 \micron \mg grains in the Rayleigh regime simultaneously account for the steep scattering slope observed by NIRISS and generate a broad silicate feature consistent with the MIRI observations.
We therefore propose that \mg can serve as a proxy for silicate clouds composed of mixtures such as SiO and SiO$_2$ \citep{dyrek2024}, and that a proper treatment of such mixed silicate compositions would likely reproduce the MIRI feature more faithfully.
Taken together, these results point towards a multi-modal cloud particle size distribution: a small mixing ratio of sub-micron grains ($\sim$0.1 \micron) is required to produce the Rayleigh scattering slope and the MIRI silicate feature. In contrast, the dominant mixing ratio of micron-sized particles ($\sim$1--5 \micron) governs the cloud deck structure and sets the overall level of spectral muting seen in NIRCam.
Below, we describe and compare their global cloud distributions, radiative feedback, and dynamical responses within a unified framework.

All three cloud species produce spatially heterogeneous distributions shaped by the interplay between local thermal structure, large-scale circulation, and particle settling. In every case, the lower boundary of the cloud deck is controlled by the condensation level set by the temperature structure. At the same time, the vertical extent depends on the balance between vertical mixing and gravitational sedimentation.

In the 5 \micron \nas case, the cloud deck extends globally to pressures lower than 0.1 bar. The cloud abundance is larger at high latitudes, while the equatorial region becomes depleted above $\sim0.01$ bar, coinciding with the super-rotating equatorial jet (Fig. \ref{fig:uwna2s}). From Fig. \ref{fig:spec_best3} (panel a), a clear longitudinal asymmetry is also present, with reduced cloud opacity on the dayside and stronger condensation on the nightside.
Because of their large optical depth, these clouds produce strong radiative feedback, leading to a net greenhouse effect. On the dayside, temperatures increase at pressures where the cloud opacity is high, consistent with absorption of outgoing thermal radiation. The nightside, in contrast, experiences a more uniform warming because of the enhanced greenhouse effect. This change in the heating structure modifies the atmospheric circulation (Fig. \ref{fig:uwna2s}): retrograde flow develops at high latitudes, the equatorial jet strengthens, and vertical velocities increase, indicating a stronger coupling between radiation and atmospheric dynamics.

In contrast, the 1 \micron KCl clouds are vertically confined between $\sim 10^{-2}$ and $10^{-4}$ bar. Although distributed over most of the planet, depletion occurs at high latitudes due to local thermal conditions and within a narrow equatorial band ($-10^\circ$ to $10^\circ$) associated with the super-rotating jet, where strong zonal winds inhibit cloud accumulation through mixing and longitudinal advection (Fig. \ref{fig:uwkcl}). Limb asymmetries arise from the interaction between the equatorial jet and longitudinal temperature contrasts, enhancing cloud abundance on the evening limb relative to the morning limb (Fig. \ref{fig:spec_best3}: panel b). Despite reproducing the transmission spectrum, the radiative impact of KCl clouds on the global circulation is modest. These clouds lead to a net warming of the atmosphere. The primary dynamical response is the emergence of retrograde winds at low pressures in polar regions, while the equatorial jet remains largely unchanged in strength and vertical extent.

The 5 \micron \mg model is characterized by efficient gravitational settling, with most condensate residing in the deep atmosphere. Nevertheless, a fraction of the cloud mass remains vertically extended to pressures sufficiently low to affect transmission. This model produces strong radiative feedback, globally modifying the thermal structure by warming the atmosphere. Atmospheric layers between $10^{-1}$ and $10^{-3}$ bar experience heating associated with cloud opacity, while deeper layers cool slightly due to radiative adjustment to the overlying heated region. The resulting redistribution of energy alters the vertical temperature gradient and substantially impacts the circulation. The equatorial super-rotating jet weakens and becomes vertically confined between $10^{-2}$ and $10^{-3}$ bar, and the flow outside the equatorial region transitions from predominantly prograde in the cloudless case to largely retrograde (Fig. \ref{fig:uwsil}). 

The silicate absorption feature detected in the MIRI bandpass can be broadly reproduced by 0.1 \micron \mg cloud particles, suggesting that analogous features arising from other silicate species at the same particle size will show similar cloud behavior with a better transmission feature match between 8 - 10 \micron. These clouds are distributed nearly homogeneously throughout the atmosphere, with a mild depletion near the equator attributable to the equatorial jet. The cloud deck reflects incoming stellar radiation, leading to reduced atmospheric temperature, a strong equatorial jet, and significant vertical wind velocities on the dayside (Fig. \ref{fig:uwsil}).

Overall, we identify three distinct scenarios that are consistent with the WASP-107b NIRCam transit data. These three scenarios have very different total cloud abundance, optical properties, and spatial distribution. To distinguish them, we can look at their emission spectra, which are close to a blackbody (Fig. \ref{fig:spec_best3}: panel d). However, differences in cloud radiative properties lead to distinct modifications of the dayside thermal structure. Although moderate, these temperature differences help break degeneracies between cloud compositions that appear similar in transmission. This demonstrates the importance of fully coupled cloud–radiation–dynamics simulations. Cloud-coupled GCMs provide a self-consistent global framework that links transmission and emission constraints and enables a systematically motivated interpretation of atmospheric structure.

\subsection{WASP-107b vs WASP-80b: Comparative Cloud–Radiation–Dynamics Behavior}
\label{section:107vs80}
Two cloud models, 1 \micron KCl and 5 \micron \mg, successfully reproduce the transmission and emission spectra of WASP-80b and the transmission spectrum of WASP-107b. In Fig. \ref{fig:compare107}, we compare how the same cloud species and particle sizes behave within the GCM framework under differing planetary parameters, with particular emphasis on temperature, gravity, metallicity, and circulation.

\subsubsection{1 \micron KCl}

\begin{figure*}[!hbt]
\centering
\includegraphics[width=\textwidth]{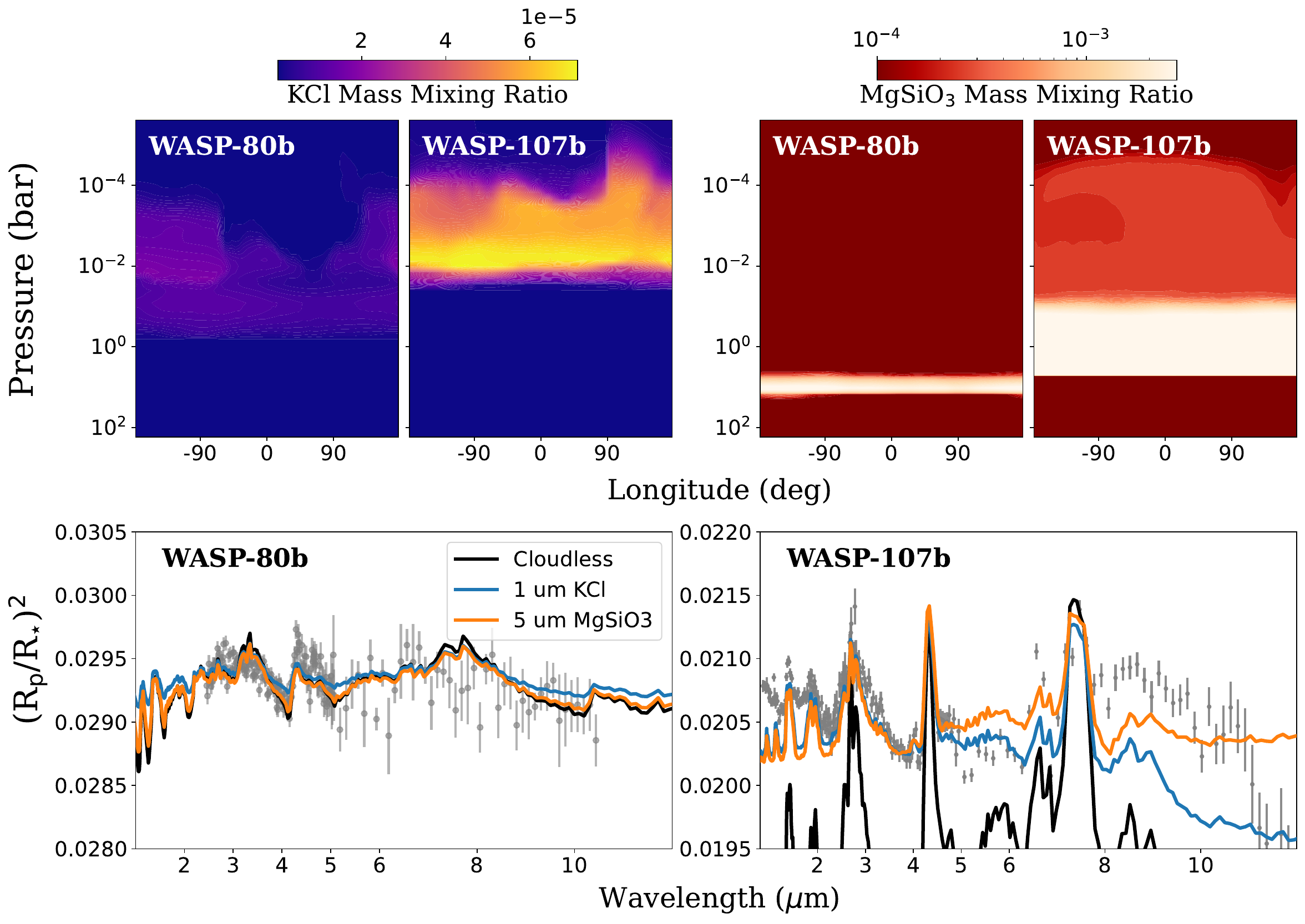}
\caption{Comparison of cloud distributions and transmission spectra for WASP-80b and WASP-107b. Top row: Longitudinal distribution of \nas clouds (shows dayside and nightside of the planet), averaged over latitudes. Within each subplot, the x-axis shows the latitude and the y-axis the pressure. Bottom row: Transmission spectra with JWST observations in gray.
}
\label{fig:compare107}
\end{figure*}

From Fig. \ref{fig:compare107}, the 1 \micron KCl case, the spectral impact differs between the two planets. In WASP-80b, these clouds have no significant effect on either the transmission or emission spectra. In contrast, for WASP-107b, 1 \micron KCl clouds strongly mute the transmission features, enabling agreement with the observations. This difference arises primarily from temperature constraining the cloud abundance and gravity constraining the vertical distribution.

The thermal structures of the two planets differ in the upper atmosphere. The dayside and evening limb of WASP-80b are hotter at low pressures compared to WASP-107b, while temperature variations on the morning and nightside are comparatively minor. At higher temperatures in WASP-80b, evaporation reduces cloud abundance, whereas the cooler atmosphere of WASP-107b favors condensation and therefore higher cloud abundance.

The lower cloud boundary differs between the two planets and is set by the combined effects of temperature and gravity. For WASP-107b, this boundary occurs at higher altitudes due to the cooler atmospheric structure. Both planets exhibit limb asymmetries, although with opposite behavior due to the temperature differences: clouds are more abundant on the morning limb of WASP-80b, whereas they are enhanced on the evening limb of WASP-107b. Dynamically, however, the wind structures remain broadly similar in the 1 \micron KCl case, with only minor changes in wind strength.

\subsubsection{5 \micron \mg}

The behavior of 5 \micron \mg clouds shows similar systematic trends but with stronger radiative feedback. As in the KCl case, these clouds do not significantly alter the transmission or emission spectra of WASP-80b, yet they strongly mute the transmission features of WASP-107b. Again, temperature primarily controls abundance, while gravity determines vertical distribution.

Silicate clouds are optically thick and therefore produce substantial radiative feedback. Differences in metallicity and cloud abundance amplify these effects in WASP-107b. Higher metallicity leads to a greater condensate mass, thereby increasing atmospheric heating throughout the planet. Because clouds in WASP-107b are present at lower pressures, they more effectively raise temperatures in these regions than in WASP-80b, where most condensate settles deeper.

Dynamically, the two planets exhibit more pronounced differences in the 5 \micron \mg model (Fig. \ref{fig:uwsil}). WASP-107b develops a smaller and weaker equatorial jet accompanied by retrograde winds, whereas WASP-80b maintains a stronger and more vertically extended jet. In both planets, most of the silicate condensate resides deep in the atmosphere due to settling; however, a small fraction remains aloft at low pressures, sufficient to influence the transmission spectrum. In WASP-107b, the weakened equatorial jet contributes to cloud depletion near the equator, while in WASP-80b, the dominant effect is gravitational settling to deeper layers.

The behavior of clouds on different planets can be explained by differences in surface gravity. 
Gravity governs the efficiency of gravitational settling and therefore controls the vertical extent of the cloud layer. 
WASP-80b has $\sim 4\times$ the gravity of WASP-107b, resulting in more efficient particle settling and cloud decks located deeper in the atmosphere. 
As a consequence, WASP-80b exhibits comparatively cloudless spectra \citep{mehta2026}. 
In contrast, the lower gravity of WASP-107b allows clouds to remain lofted to lower pressures, producing a more vertically extended cloud distribution. 

These comparisons demonstrate that identical cloud species and particle sizes can produce substantially different observable signatures depending on planetary temperature and gravity. 
Lower gravity and cooler temperatures favor vertically extended clouds and muted transmission spectra, whereas higher gravity and hotter atmospheres promote deeper cloud decks and reduced spectral impact. 
In addition, metallicity can shift condensation curves by altering the thermochemical stability of condensates, changing the pressure levels at which cloud formation occurs, and further modulating the resulting cloud opacity structure. 
This highlights the importance of coupling clouds, radiation, and dynamics within a self-consistent GCM framework when interpreting comparative exoplanet atmospheres.

\subsection{WASP-107b: Limb Asymmetry}
\label{section:la}

\begin{figure*}[!hbt]
\centering
\includegraphics[width=\textwidth]{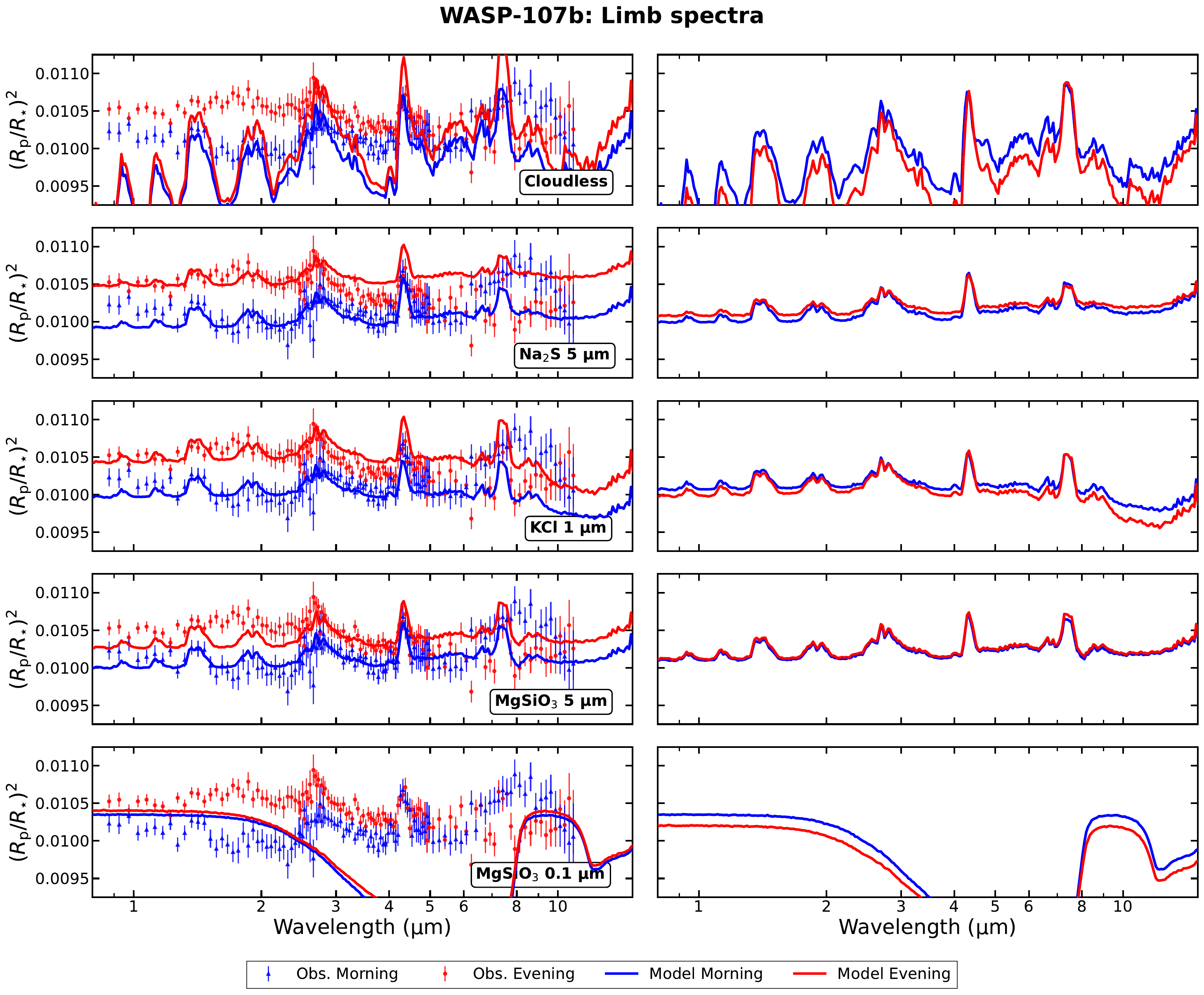}
\caption{Morning and evening limb spectra for the best-fit models of WASP-107b, including the 0.1 \micron \mg cloud case, compared with JWST observations from \cite{murphy2025}. Left column: Transmission spectra from both limbs as obtained from the GCM, plotted alongside the observed data. Right column: Spectra aligned at the CO$_2$ absorption feature near 4.3 \micron to highlight the limb-to-limb differences induced by cloud asymmetry.}
\label{fig:best3la}
\end{figure*}

WASP-107b exhibits a pronounced limb asymmetry in the JWST observations \citep{murphy2024, murphy2025}. Among the three cloud cases that reproduce the transmission spectrum, we evaluate their ability to explain the observed differences between the morning and evening limb spectra. 
Fig. \ref{fig:best3la} presents the spectra of the best-fit cloud models, 0.1 \micron \mg together with the cloudless case, and the resulting spectra from the morning and evening limb. 

We analyze the limb asymmetry using two complementary approaches. In the first (Fig. \ref{fig:best3la}: Left column), we compute the transmission spectrum for each limb and compare it to the JWST observations, accounting for both the temperature difference and the cloud deck altitude at each limb. 
In the second (Fig. \ref{fig:best3la}: Right column), we align the CO$_2$ peak of each limb spectrum and compare the relative feature strengths, isolating the effect of temperature structure and cloud opacity on the spectra independently of any absolute depth offset.

Both approaches carry their own assumptions, the validity of which depends on how the individual limb spectra are extracted from the observations. 
The morning limb dominates the ingress signal, and the evening limb the egress, but the transition is gradual. The precise partitioning depends on the adopted contact times and on the treatment of the overlapping mid-transit region. In practice, this transition is difficult to resolve even with JWST, and light-curve fits typically assume a sharp, unphysical evening--morning boundary to simplify the modeling \citep{espinoza2021}. 
Any systematic uncertainty in this partitioning propagates directly into the inferred inter-limb depth offset, making the first approach sensitive to these timing choices. 
The second approach is largely immune to such systematics, as aligning the CO$_2$ peak removes any common-mode offset and retains only the differential spectral shape. 
Where both approaches yield consistent conclusions, the result can be considered robust; where they diverge, the relative CO$_2$ feature strength serves as the more conservative diagnostic. For the WASP-107b data used here, the transit timing was derived by fitting several multi-wavelength, multi-epoch datasets simultaneously \citep{murphy2024}, which is considered best practice and should yield a robust measurement of the limb-to-limb offset \citep{powell2019}. Therefore, our first approach should provide a near-exact comparison. We nonetheless show the models aligned with the CO$_2$ feature in the left column of Fig. \ref{fig:best3la} to facilitate comparison with future work. 

The cloudless model establishes the baseline for comparison: it produces a peak limb temperature difference of approximately 200 K at $\sim10^{-5}$ bar (Fig. \ref{fig:best3}), consistent with the temperature difference observed in \cite{murphy2024}. Still, the spectra are incompatible with the observations as clouds are required on both limbs to reproduce the observed transit depths. The offset within the molecular bands is consistent with the observations; however, outside these bands, deeper atmospheric layers are probed, where the temperature contrast is reduced.
Furthermore, the right column of Fig. \ref{fig:best3la} shows that the CO$_2$ feature is stronger on the evening limb than on the morning limb, which is inconsistent with the observed trend.

When clouds are included in the models, a large offset is observed between the two limb spectra. This is not due to a significant change in the limb-to-limb temperature difference when clouds are added, but rather to the fact that this limb-to-limb temperature difference increases with decreasing pressure. In the cloudless case, limb asymmetry is primarily confined to the bands that probe the lowest pressures. When clouds are added in the model, the deeper atmosphere, where no asymmetry exists, is not probed anymore, and only the parts where a strong limb-to-limb temperature gradient is probed, leading to approximately constant offsets between the two limbs. Although the three cloud scenarios that fit the averaged spectrum have a similar effect on the limb spectra, some differences remain. 

In the 5 \micron \nas the peak temperature difference is reduced to $\sim$180 K and shifts to $10^{-3}$-$10^{-2}$ bar (Fig. \ref{fig:best3}). Whereas the morning spectra is a good match to the data shorter than $7 \mu m$, the evening spectra overestimates the cloud effect in the main window between the CH$_4$ and the CO$_2$ band. In contrast, the 1 \micron \kcl case does not affect the limb temperature gradient, but the match to the observations is better in the $3-4\micron$ case than for the \nas case. The  5 \micron \mg clouds produce a markedly different temperature structure: the contrast is broadly distributed between $10^{-4}$ and $10^{-1}$ bar, peaking at $\sim$150 K, with an additional contrast at very low pressures ($\sim10^{-4}$ bar). Whereas the 3-4 \micron window of the evening spectra is better matched, the spectra fall below the observations at shorter wavelengths. 

Another comparison is the strength of the CO$_2$ feature in the evening and morning spectra. As can be seen in the right panel of Fig \ref{fig:best3la}, the addition of clouds makes the strength of the CO$_2$ feature very similar between the two limbs, whereas it is smaller in the evening limb in the observations. Only the 5 \micron \nas cloud model conserves a stronger evening CO$_2$ feature. 

Finally, we note that none of the models reproduce the morning spectra beyond 7 \micron. Because the large feature was attributed to silicate clouds in previous work, we also present the 0.1 \micron \mg model for completeness. The model indeed produces a large silicate feature. However, because 0.1 \micron \mg particles are well homogenized over the whole planet, they do not lead to any limb difference in that wavelength range. Overall, our models show that small silicate particles are very unlikely to exhibit limb-to-limb asymmetry, which calls into question the interpretation of this spectral feature as arising from silicate clouds. 

Alternatively, a multi-modal size distribution, in which a small mixing ratio of sub-micron \mg grains produces the MIRI silicate feature, could naturally give rise to limb asymmetries if the abundance or size distribution of this sub-micron population differs between the evening and morning terminators. Assessing whether such a scenario is likely would require complex microphysics models coupled to the circulation.

There may also be unrealized biases present in the MIRI limb spectra that affect the relative evening--morning depths.
Emission spectroscopy of WASP-107b, particularly with MIRI, would help resolve this discrepancy by providing independent constraints on the dayside chemical and cloud properties that feed the evening limb.

\subsection{WASP-69b: Comparison with Observations}

\begin{figure*}[!hbt]
\centering
\includegraphics[width=\textwidth, trim=10 10 5 10, clip]{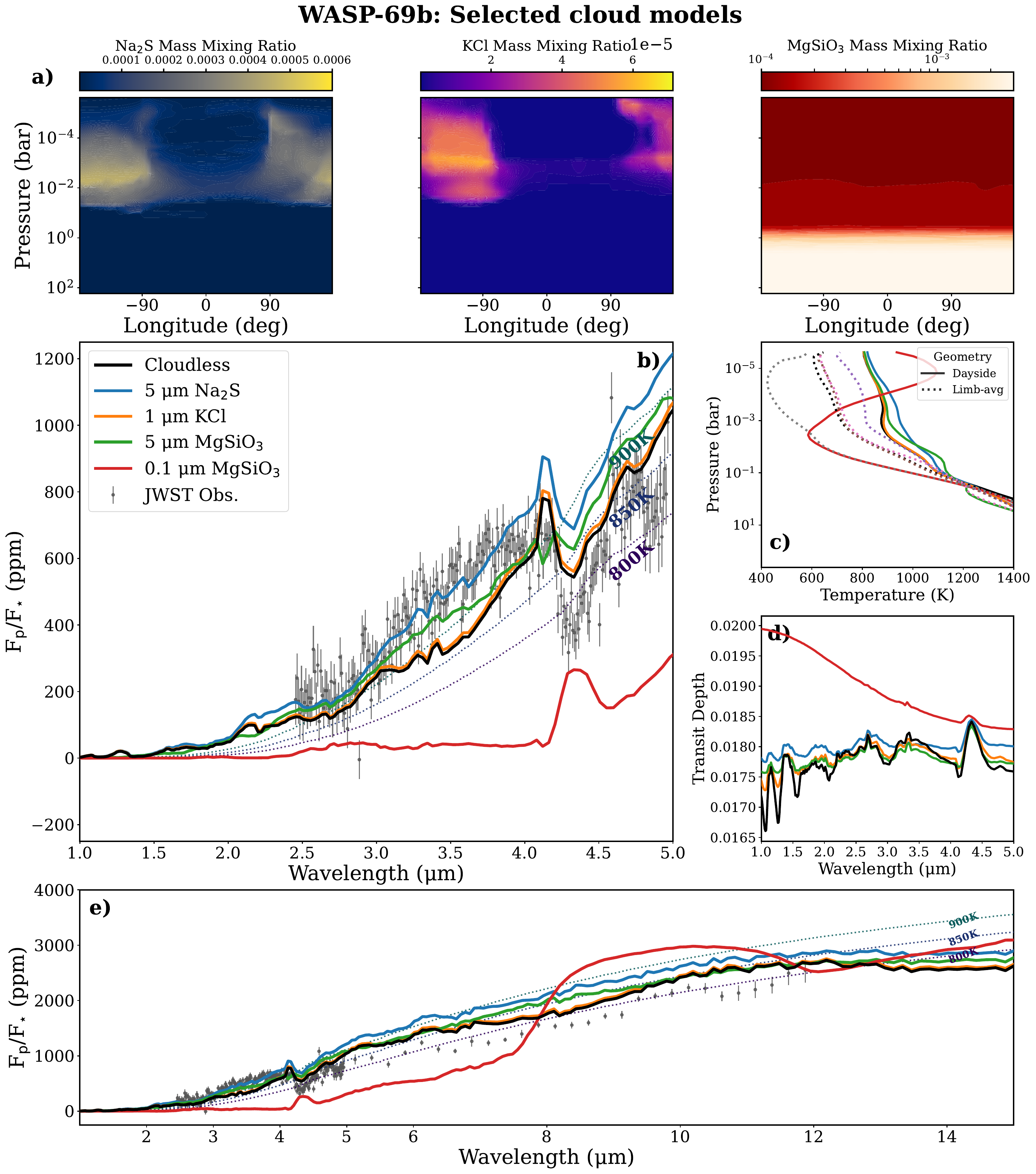}
\caption{Panel \textbf{a} shows the longitudinal distribution of Na$_2$S (left), KCl (middle), and MgSiO$_3$ (right) clouds (showing dayside and nightside), averaged over latitudes for WASP-69b models. Panel \textbf{b} shows the emission spectra of the 3 models (best-fit Wasp-107b models) along with the JWST observation for NIRCam. Panel \textbf{c} shows corresponding limb-averaged (dotted) and dayside averaged (solid) TP profiles with emission spectra in panel \textbf{d}. Panel \textbf{e} shows NIRCam + MIRI observations with the models.}
\label{fig:spec_best3_69}
\end{figure*}

We compare the model spectra with the available observations across the F322W2, F444W, and MIRI wavelength regions. 
Fig. \ref{fig:spec_best3_69} shows the results for WASP-69b, applying the same set of models that successfully reproduced the spectrum of WASP-107b.
Unlike WASP-107b, none of the models can fully reproduce the observations of WASP-69b, suggesting that the atmospheric conditions of this planet require a different physical configuration.

For the Na$_2$S cases, all models reproduce the F322W2 spectral region (Fig. \ref{fig:tp69}). 
However, they fail to provide a satisfactory fit in F444W and MIRI. In particular, the models are unable to reproduce the dip near 4.3 \micron, which requires a steep temperature gradient between deep and shallow atmospheric layers. 
In the MIRI wavelength range, the models systematically overestimate the emitted flux.

For the KCl models, the emission spectra are nearly identical to the cloudless case.
The clouds reside predominantly on the nightside and do not significantly modify the dayside temperature structure, resulting in minimal impact on the emission spectrum.

For the \mg cases, particle size plays a critical role. 
Smaller particles, in the presence of a temperature inversion, begin to produce emission features associated with cloud opacity. 
The 5 \micron particles yield higher flux levels due to increased atmospheric temperatures, whereas the 10 \micron particles settle efficiently to deeper layers and therefore have negligible spectral impact.

The influence of clouds on the transmission spectrum follows the expected trend with particle size and vertical distribution. 
As particle size increases, the baseline of the transmission spectrum rises, reflecting the vertical settling of the cloud deck. 
The 0.1 \micron Na$_2$S case enhances the Rayleigh slope owing to the small particle size, while 0.1 \micron \mg produces a comparable slope in addition to cloud-specific spectral features.

\begin{figure*}[!hbt]
\centering
\includegraphics[width=\textwidth]{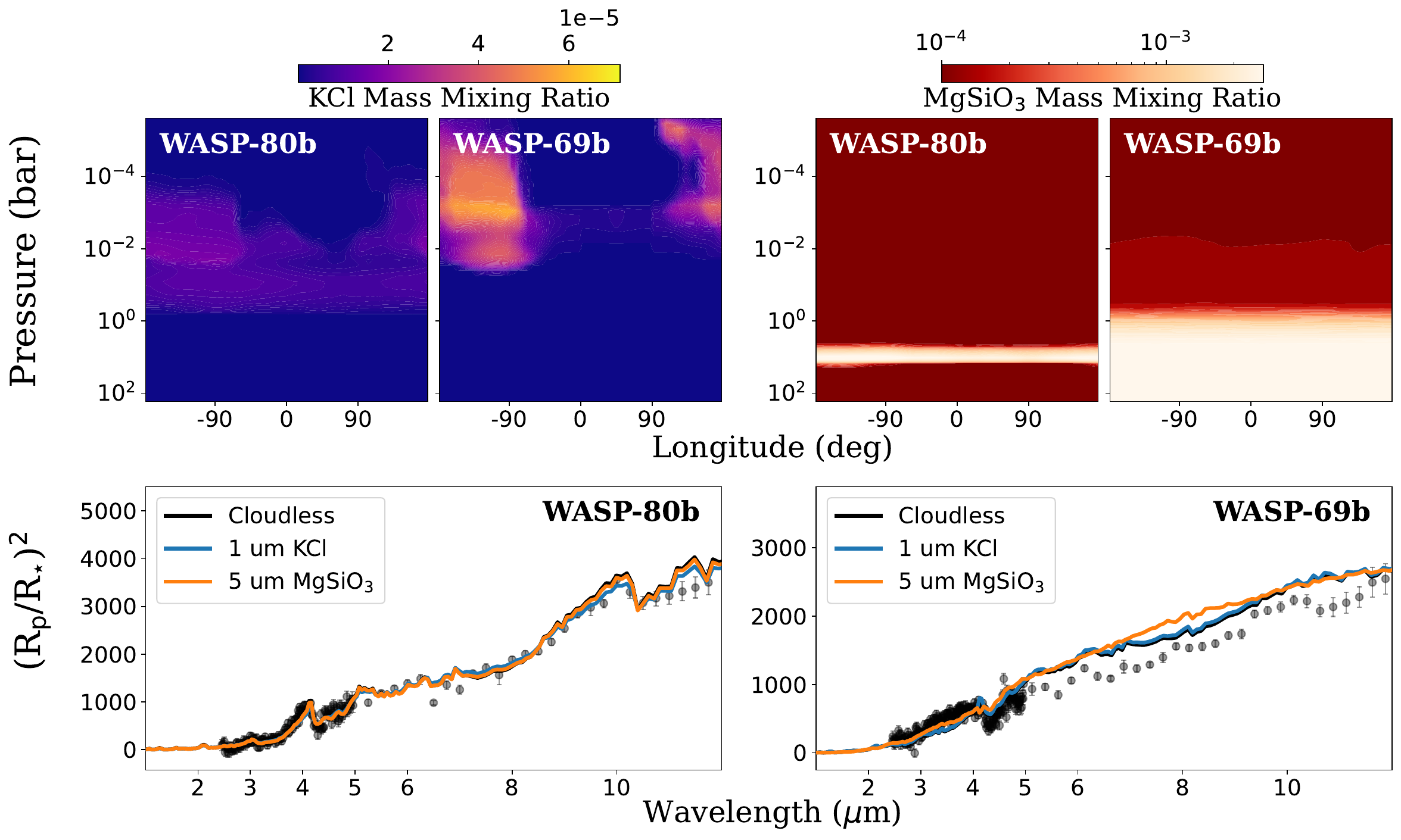}
\caption{Comparison of cloud distributions and transmission spectra for WASP-80b and WASP-69b. Top row: Longitudinal distribution of \nas clouds (shows dayside and nightside of the planet), averaged over latitudes. Within each subplot, the x-axis shows the latitude and the y-axis the pressure. Bottom row: Transmission spectra with JWST observations in gray.
}
\label{fig:compare69}
\end{figure*}

We further compare the 1 \micron KCl and 5 \micron \mg models against the WASP-80b results (analogous to Fig. \ref{fig:spec_best3} and Section \ref{section:107vs80}, but applied to WASP-69b). For the 1 \micron KCl case, the dayside clouds are fully evaporated due to the higher dayside temperatures, while the cooler nightside supports a greater cloud abundance through condensation. KCl clouds in this regime carry low opacity and weak radiative feedback, and consequently produce negligible impact on the emission spectrum of both WASP-69b and WASP-80b.

The 5 \micron \mg clouds present a more nuanced picture. 
On WASP-80b, the higher gravity causes these particles to settle efficiently to depth, rendering them radiatively inactive and leaving the emission spectrum largely unaffected. 
On WASP-69b, by contrast, the lower gravity allows the clouds to remain suspended at pressures as low as 0.01 bar, where their higher opacity raises the atmospheric temperature and increases the emitted flux at wavelengths shortward of 4 \micron, bringing the models into closer agreement with the observations in that region. 
Nevertheless, the 4.3 \micron absorption feature remains poorly reproduced in both models, which are unable to match the MIRI observations, indicating that the atmospheric structure of WASP-69b poses challenges beyond what the current cloud configurations can address.

Overall, the cloud-coupled GCM models are unable to reproduce the emission and transmission observations in a self-consistent manner. The observations may instead require a mixture of cloud species with distinct vertical distributions: high-absorption clouds in the deep atmosphere (e.g., Na$_2$S) that enhance heating and increase flux in F322W2, combined with more reflective clouds in the upper atmosphere that reduce temperatures and modify the emission spectrum. Such a vertical cloud mixture could solve the discrepancies between wavelength regions.

Exploring mixtures of cloud species within the GCM framework, including combinations of deep Na$_2$S clouds and reflective upper-atmosphere clouds, is beyond the scope of this paper and is left to future work. This can be implemented using ADAM, although the computational cost of exploring the full parameter space of cloud mixtures is substantial.

\section{Conclusion}
\label{section:conclusion}

We have presented a systematic GCM study of three warm Jupiters, WASP-107b,
WASP-80b and WASP-69b, spanning a range of planetary properties, and compared
model outputs directly against JWST observations. Our key findings are as follows.

\begin{enumerate}

\item Small differences in gravity, metallicity, and temperature can lead to strong variations in spectra. These variations are most likely due to increased cloud settling on high-gravity planets.

\item Equatorial super-rotation develops in most of our models, leading to a hotter equator and a higher cloud base. Furthermore, the circulation transports particles out of the jet, reducing the cloud top. Both lead to a lower overall cloud abundance at the equator.

\item Clouds are further shaped by the day-night contrast difference, with a depletion of \nas and \kcl clouds on the dayside for all three planets, whereas \mg clouds are more homogeneously distributed. 

\item Current observations of WASP-107b favor 5 \micron \nas, 1 \micron KCl, and 5 \micron \mg clouds. Clouds formed of smaller particle sizes remain lofted and produce strong Rayleigh scattering clouds, whereas clouds with larger particles settle too deep in the atmosphere to affect the spectra. 

\item The WASP-107b cloud models applied to WASP-80b do not change the optical thickness. The greater gravity of WASP-80b leads to deeper, more compact clouds than those of WASP-107b. This shows that gravity is a key parameter for understanding the spectral diversity of warm Jupiters.

\item The shift between the evening and morning limb measurements of WASP-107b can be well reproduced by the three cloud models fitting the averaged spectra. The change in spectral shape between the two limb spectra is, however, not reproduced correctly. None of the models can match the change in the size of the CO$_2$ absorption feature. The \nas cloud is the only one producing a larger feature in the evening limb, but that difference is not sufficient to explain the observations. 

\item We are unable to reproduce the WASP-107b MIRI spectrum, particularly the large absorption band at 8-10 \micron. Additionally, our models do not predict any limb asymmetry in a possible silicate feature due to small silicate particles. That is because small silicate clouds do not have a horizontally varying distribution. 
This raises the question of whether the feature is due to silicate clouds. 

\item The different cloud scenarios proposed for WASP-107b could be distinguished by observing the emission spectra of the planet, as the radiative feedback effect of each of them is very different. 

\item Sub-micron Na$_2$S and MgSiO$_3$ drive thermal inversions through efficient shortwave absorption, heating upper atmospheric layers while cooling deeper regions (an anti-greenhouse effect). KCl clouds exert minimal radiative impact owing to their comparatively low opacity.

\item Our models fail to reproduce the emission spectra of WASP-69b. In particular, the deep CO$_2$ feature observed in the spectrum indicates a stronger vertical temperature gradient on the dayside than predicted. Furthermore, our models do not lead to cloud and temperature distributions that correspond to the 2 temperature-pressure profile fit proposed in \citep{schlawin2024}, indicating the need to incorporate additional physical mechanisms into the models. 

\end{enumerate}

Taken together, these results demonstrate the complex and tightly coupled nature of clouds, radiation, and atmospheric dynamics in warm Jupiters. Variations in particle size, composition, and planetary properties produce diverse thermal structures and observable spectra, often with significant degeneracies that can only be broken by multi-wavelength and multi-phase constraints. Self-consistent 3D modeling, combined with complementary transmission and emission observations, can further help break these degeneracies by distinguishing among cloud scenarios that are degenerate with 1D retrievals. 

\begin{acknowledgements}

This work was supported by the French government through the France 2030 investment plan managed by the National Research Agency (ANR), as part of the Initiative of Excellence Université Côte d'Azur under reference number ANR-15-IDEX-01. The authors are grateful to the Université Côte d'Azur's Center for High-Performance Computing (OPAL infrastructure) for providing resources and support.\\

\end{acknowledgements}

\bibliographystyle{aa}
\bibliography{main}

\begin{appendix}
\label{appendix}

\section{Opacity Mixing: Bin-then-Mix versus Mix-then-Bin}
\label{appendix:opa_mix}

An important question that arises in the context of GCMs is whether correlated-$k$ tables should be binned first and subsequently mixed, or mixed first and subsequently 
binned. This distinction becomes particularly significant when implementing active chemistry, where the atmospheric opacity evolves with the local molecular abundances and the mixing of $k$-tables must be performed on-the-fly during the simulation. To assess the validity of both approaches, we conducted a series of tests and compared the results against pre-mixed $k$-tables, which serve as the reference case.

The tests are conducted for WASP-80b \citep{mehta2026}, assuming solar metallicity and a solar C/O ratio. Three configurations are compared:

\begin{enumerate}
    \item \textbf{Pre-mixed $k$-tables} (reference case): opacity tables are taken 
    directly from \cite{lupu2021} and \cite{tan2024}, where mixing has been performed before binning at full spectral resolution.

    \item \textbf{Mix-then-bin}: the per-species k-tables are first mixed according to solar abundances at full spectral resolution using \texttt{exo\_k}, and the resulting combined table is subsequently binned into 11 spectral bands.

    \item \textbf{Bin-then-mix}: the per-species opacity tables are first individually binned into 11 spectral bands, and the binned tables are then mixed according to solar abundances.
\end{enumerate}

\begin{figure*}
    \centering
    \includegraphics[width=\columnwidth]{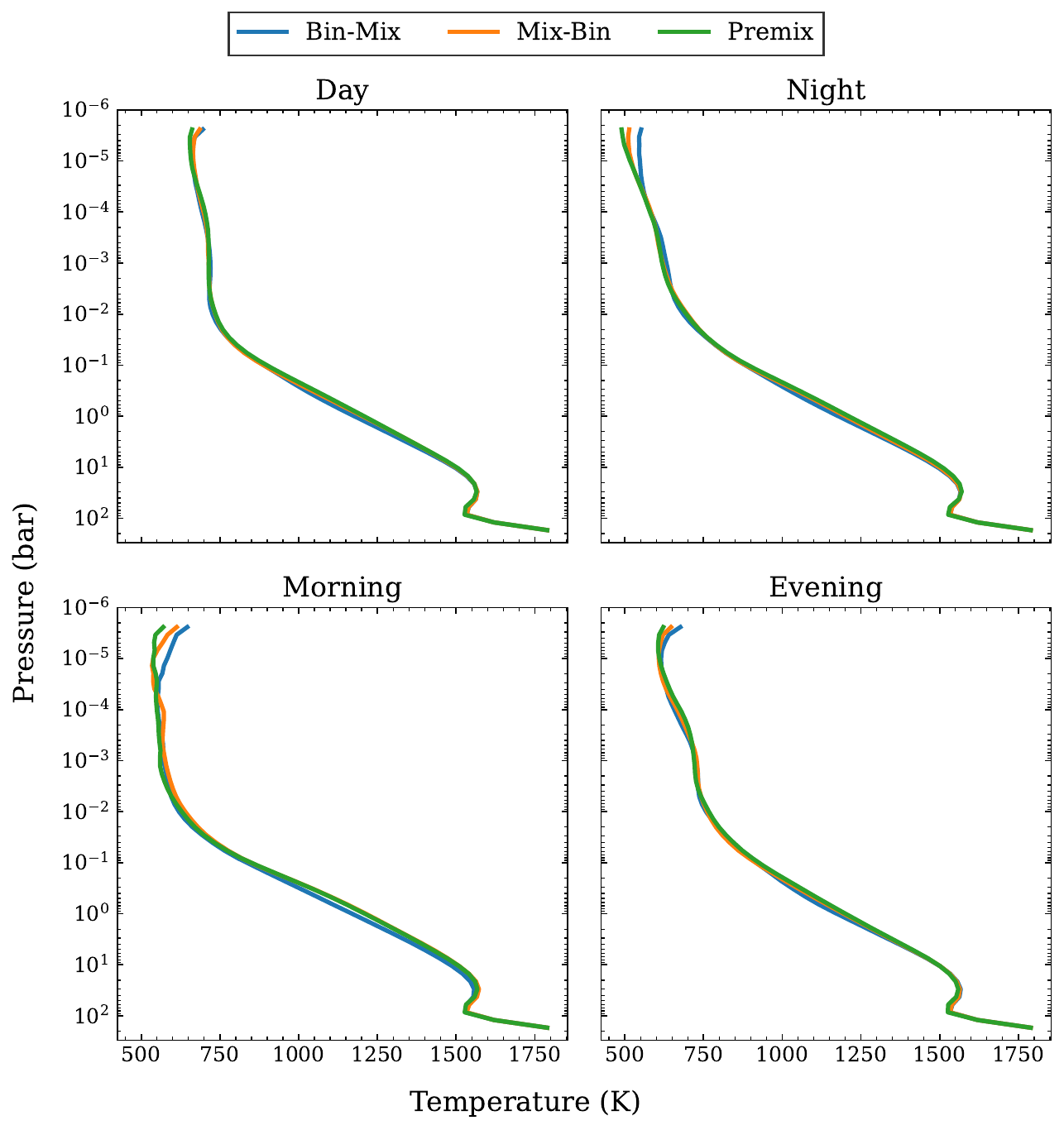}
    \includegraphics[width=\columnwidth]{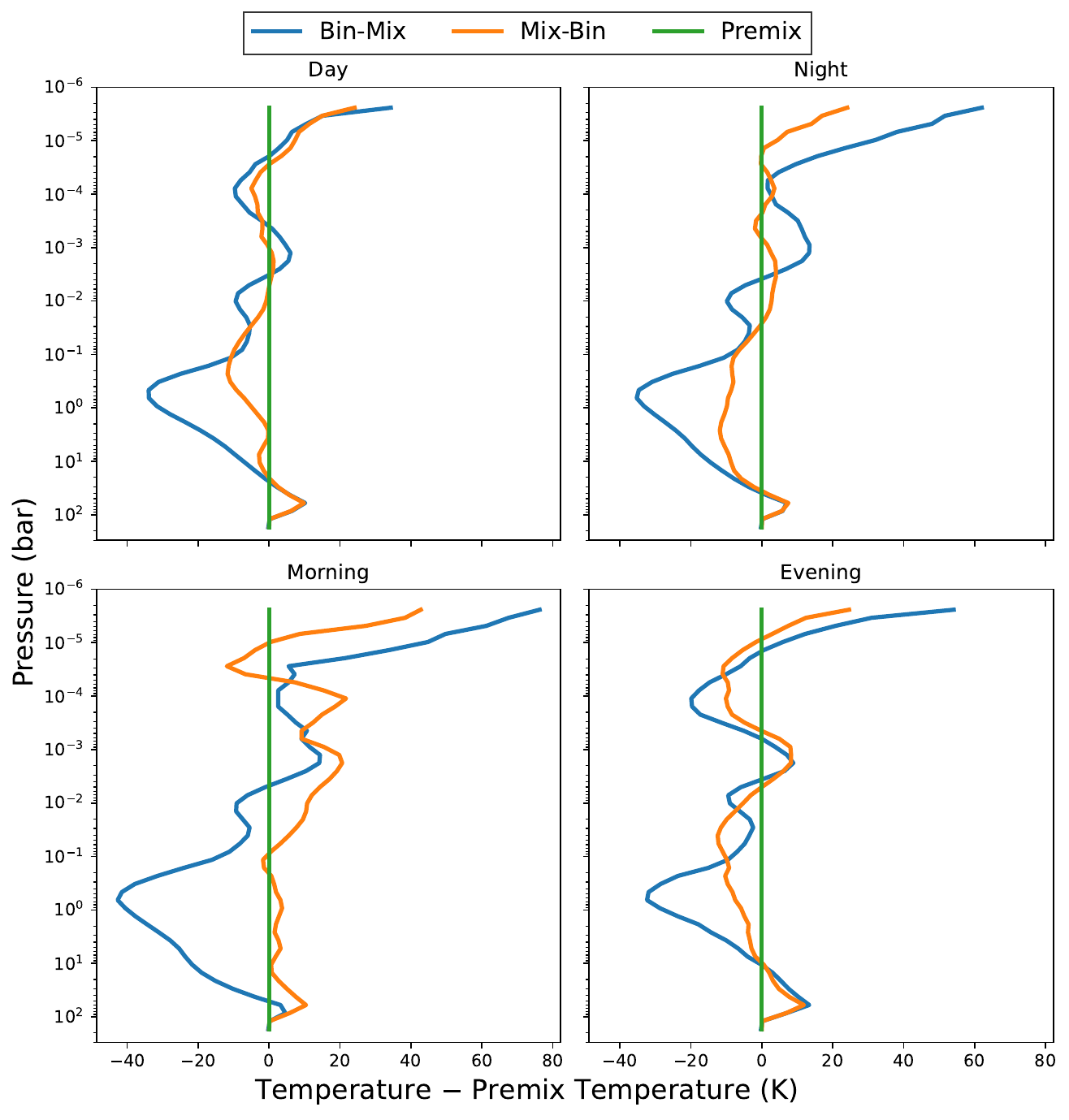}
    \caption{\textit{Left}: Temperature-pressure profiles at four regions of the planet (day, night, morning, and evening sides) for three opacity schemes: Bin-Mix, Mix-Bin, and Pre-mix. \textit{Right}: Deviation of the Bin-Mix and Mix-Bin temperature profiles from the Pre-mix reference for the same planetary regions}
    \label{fig:bm_tp}
\end{figure*}

The results show that both approaches introduce only minor differences relative to 
the pre-mixed reference, with the mix-then-bin method yielding slightly greater 
accuracy than bin-then-mix. The mean discrepancy in temperature (Fig \ref{fig:bm_tp}) amounts to $<$3\% and maximum around $<$15\% (for cooler regions), which is 
sufficiently small for practical purposes. Crucially, the bin-then-mix approach 
offers a substantial computational advantage \footnote{Benchmarks were performed using a Jupyter Notebook on an Apple M2 processor.}: mixing unbinned $k$-tables requires 
109.48 seconds, whereas mixing the pre-binned tables requires only 2.26 seconds, 
representing a 97.9\% reduction in computational cost. Here, only the mixing time 
is considered in this comparison, since the binning step occurs either before or 
after mixing and does not affect on-the-fly calculations within the GCM. This 
result demonstrates that bin-then-mix is a viable and efficient strategy for 
implementing on-the-fly chemistry with varying molecular abundances in GCMs.

\section{Supplementary information of atmospheric structure and cloud diagnostics}

\begin{figure*}[!hbt]
\centering
\includegraphics[width=\textwidth]{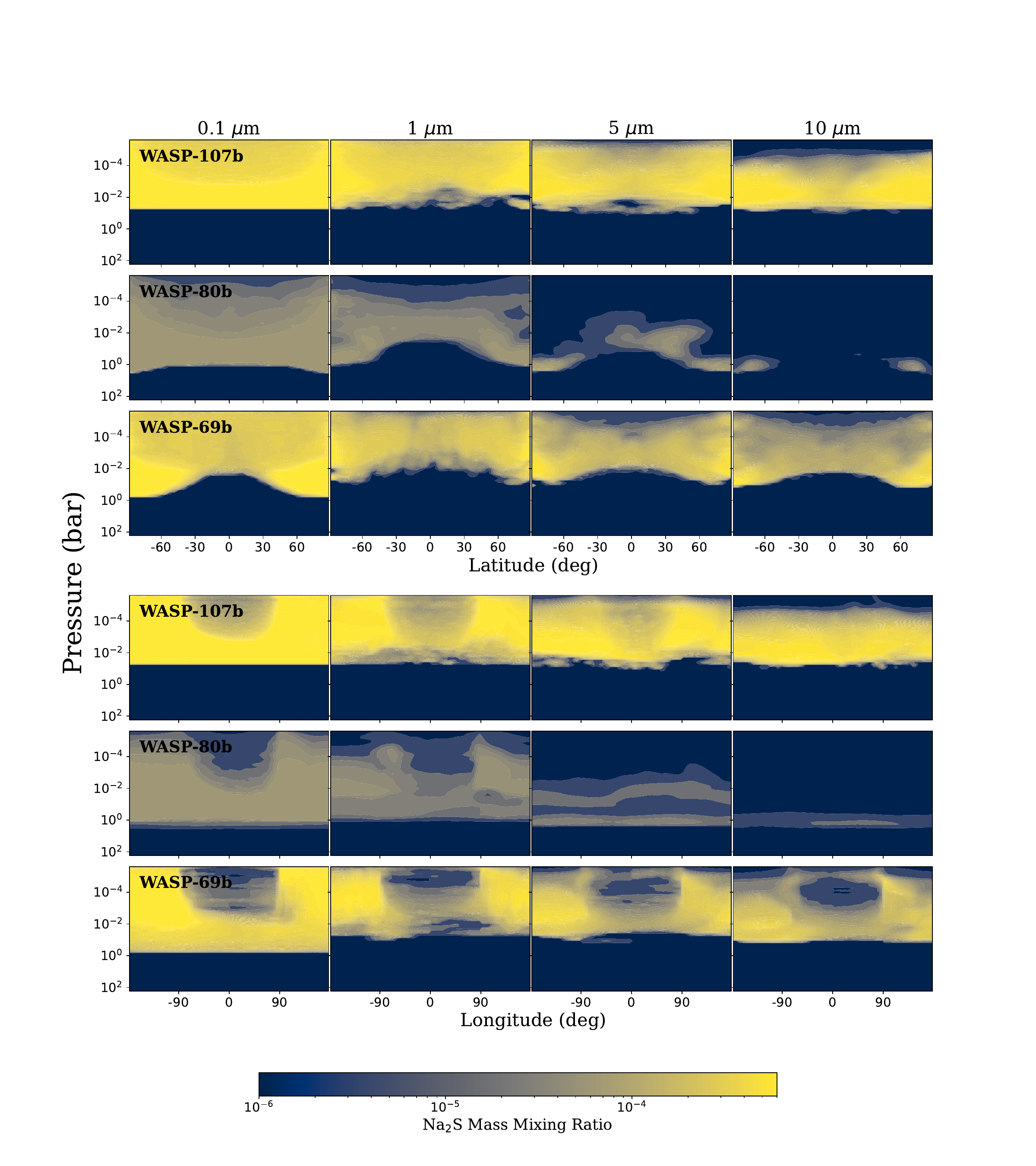}
\caption{
Top: Latitudinal distribution of \nas clouds (shows equatorial and polar regions), averaged over longitudes for all models. Within each subplot, the x-axis shows the latitude and the y-axis the pressure. Rows 1, 2, and 3 correspond to WASP-107b, WASP-80b, and WASP-69b clouds, respectively, with each column representing a different particle size indicated at the top. Bottom: Same as top, but for longitudinal distribution of clouds (shows dayside and nightside of the planet), averaged over latitudes for all models.
}
\label{fig:cldna2s}
\end{figure*}

\begin{figure*}[!hbt]
\centering
\includegraphics[width=\textwidth]{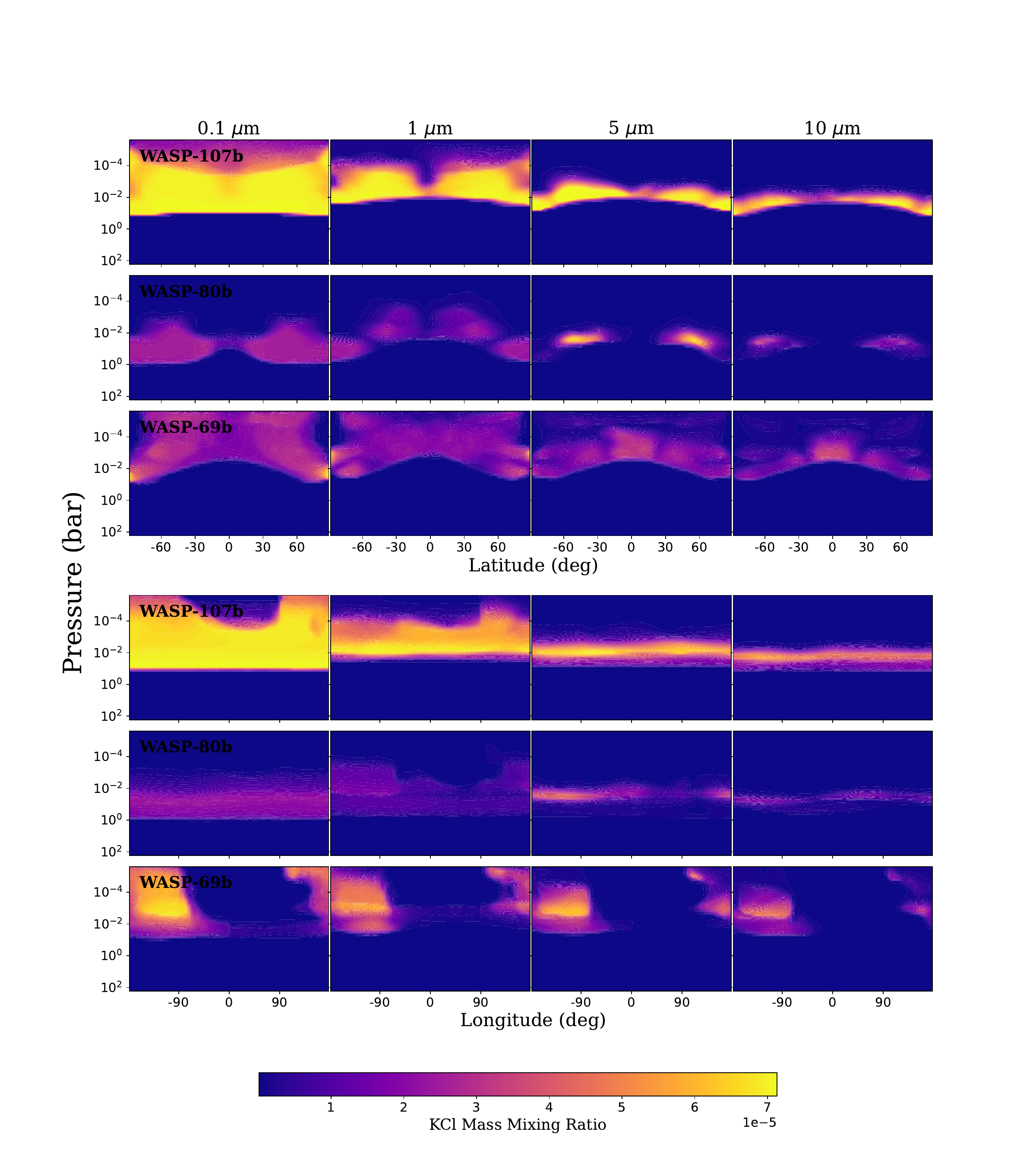}
\caption{Same as Fig. \ref{fig:cldna2s} but for KCl cloud models.
}
\label{fig:cldkcl}
\end{figure*}

\begin{figure*}[!hbt]
\centering
\includegraphics[width=\textwidth]{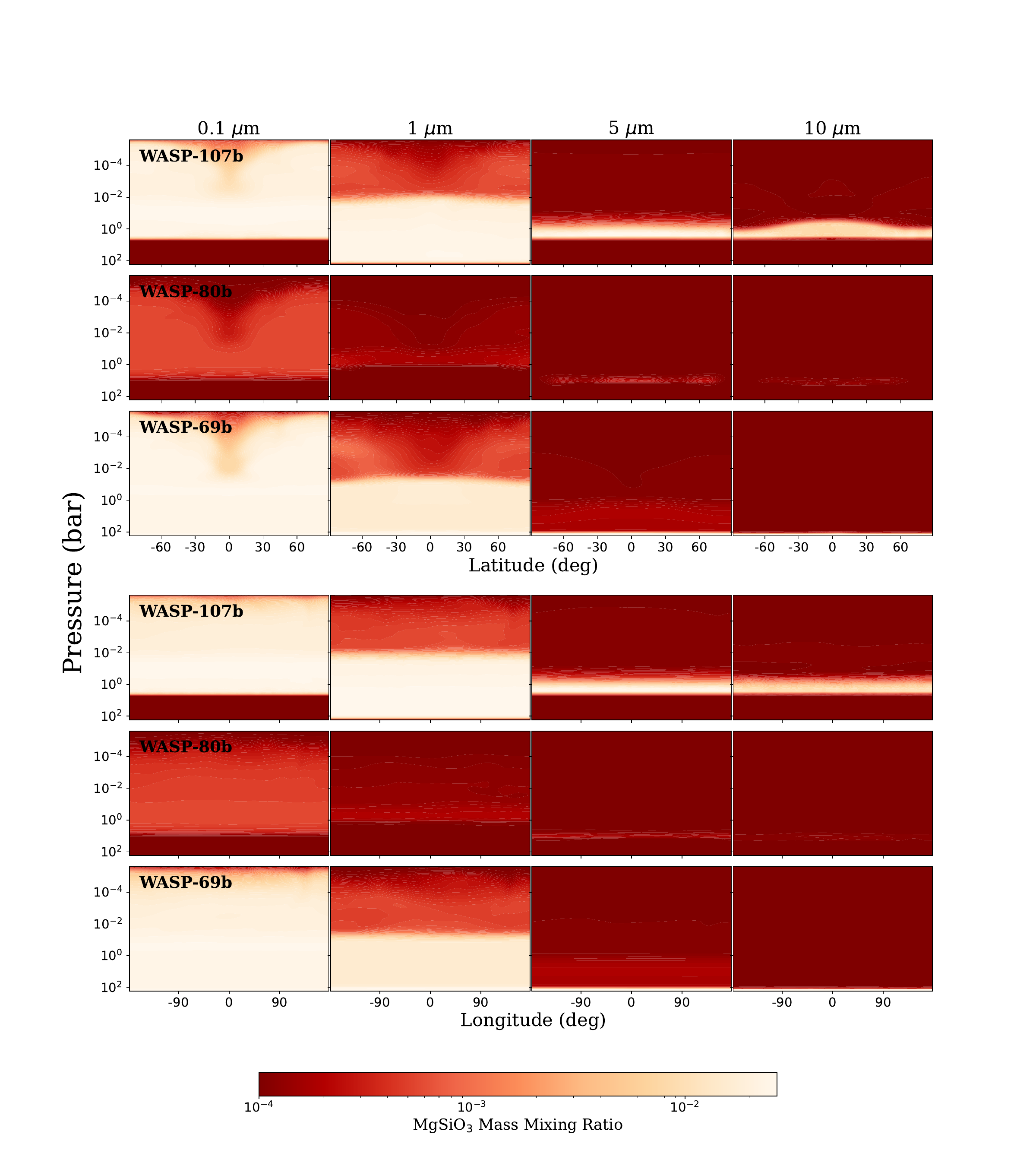}
\caption{Same as Fig. \ref{fig:cldna2s} but for MgSiO$_3$ cloud models.
}
\label{fig:cldsil}
\end{figure*}

\begin{figure*}[!hbt]
    \centering
    \includegraphics[width=\textwidth]{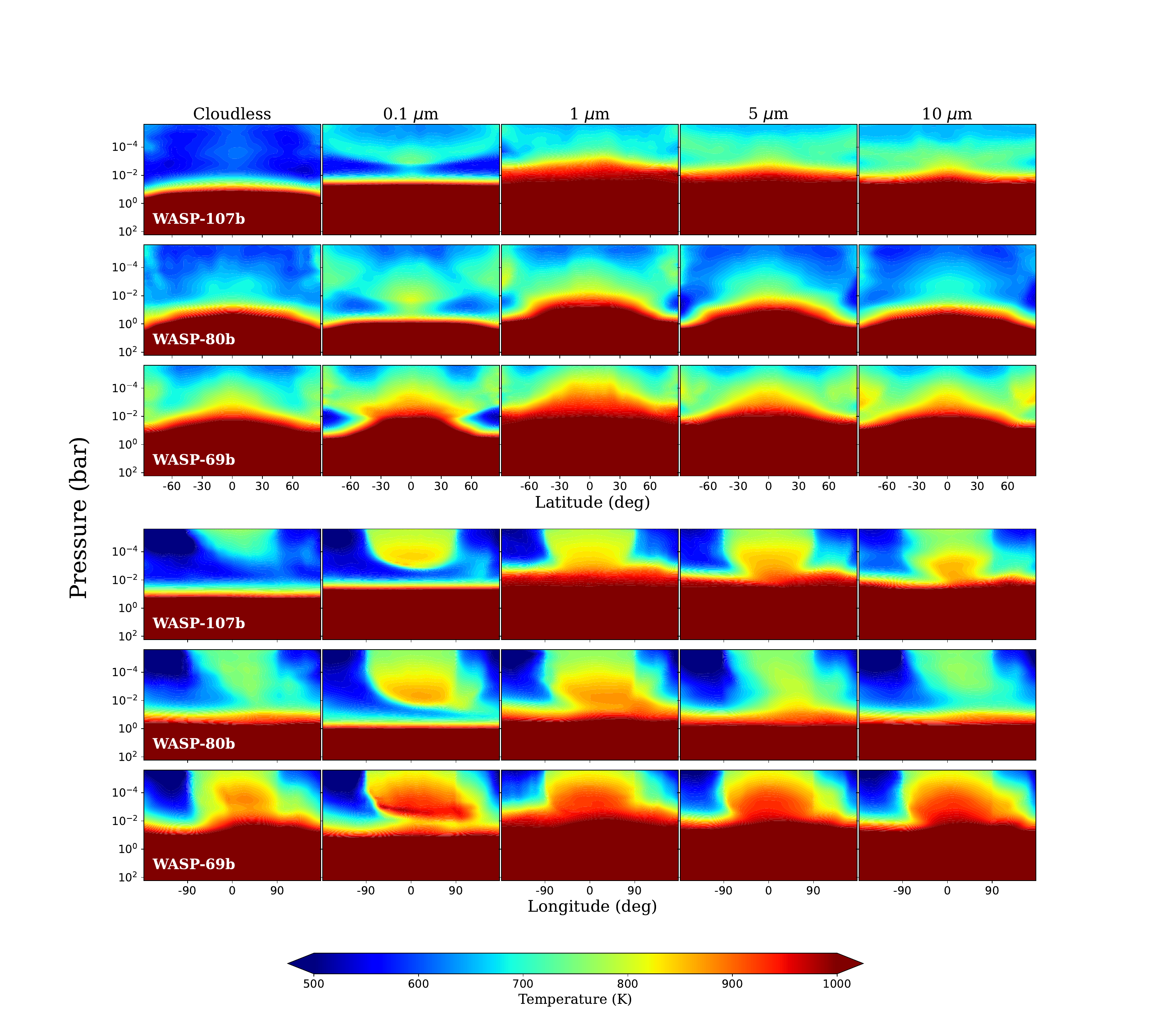}
    \caption{(\nas models) Top: Latitudinal distribution of temperature (shows equatorial and polar regions), averaged over longitudes for all models. Within each subplot, the x-axis shows the latitude and the y-axis the pressure. Rows 1, 2, and 3 correspond to WASP-107b, WASP-80b, and WASP-69b clouds, respectively, with each column representing the cloudless case and different particle sizes indicated at the top. Bottom: Same as top, but for longitudinal distribution of clouds (shows dayside and nightside of the planet), averaged over latitudes for all models.}
    \label{fig:tna2s}
\end{figure*}

\begin{figure*}
    \centering
    \includegraphics[width=\textwidth]{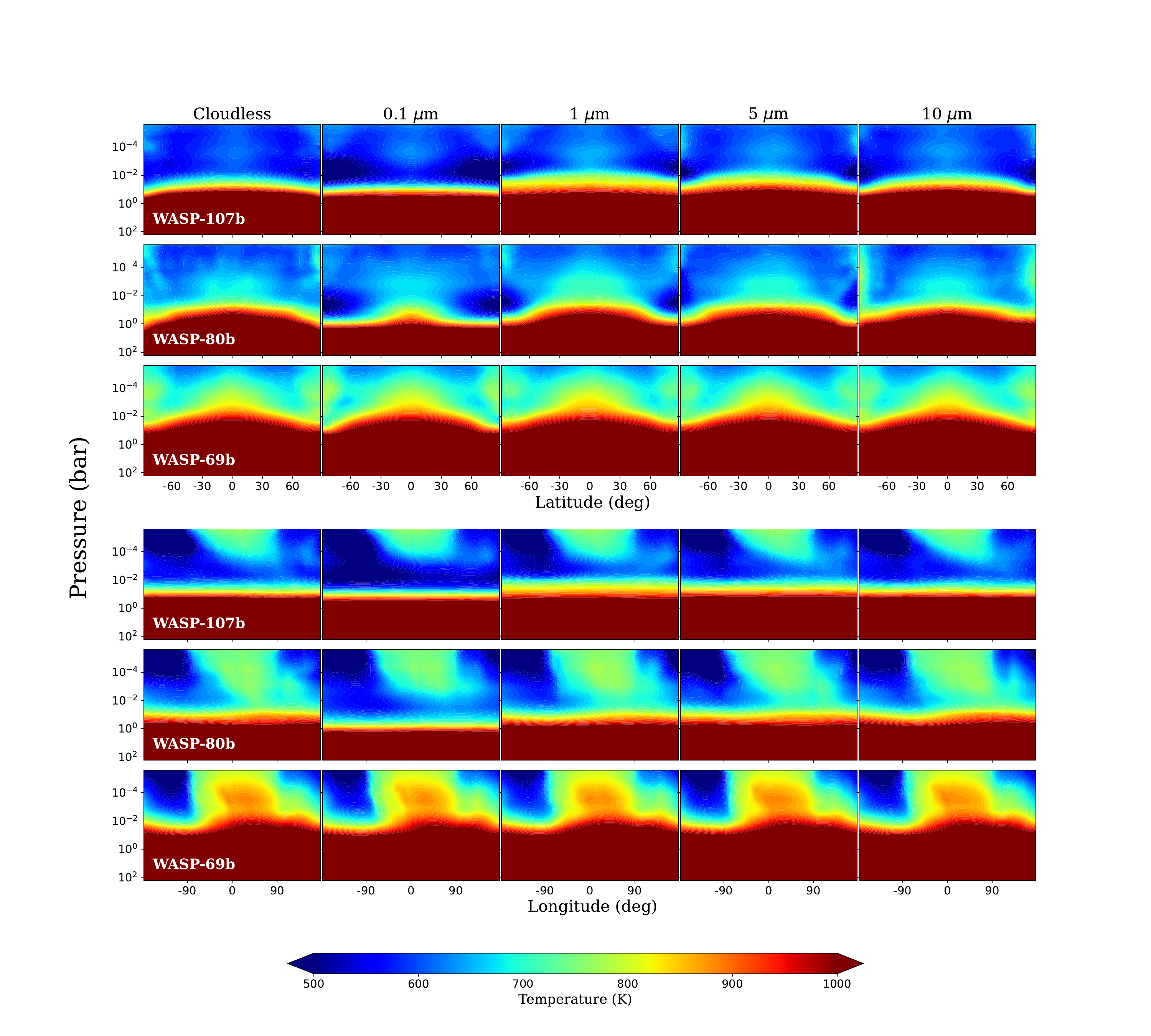}
    \caption{Same as Fig. \ref{fig:tna2s} but for KCl cloud models.}
    \label{fig:tkcl}
\end{figure*}

\begin{figure*}
    \centering
    \includegraphics[width=\textwidth]{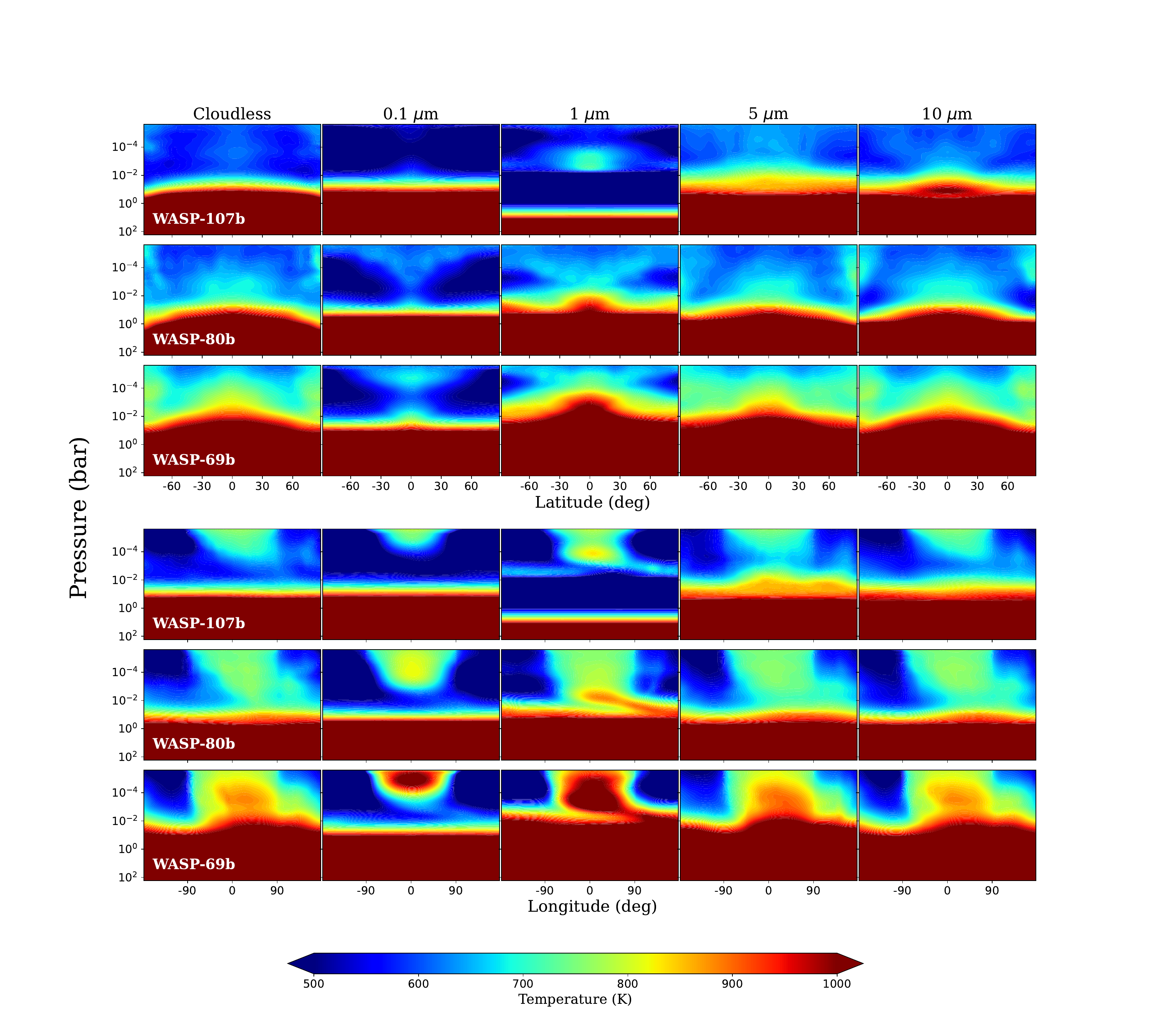}
    \caption{Same as Fig. \ref{fig:tna2s} but for \mg cloud models.}
    \label{fig:tsil}
\end{figure*}

\begin{figure*}[!hbt]
\centering
\includegraphics[width=\textwidth]{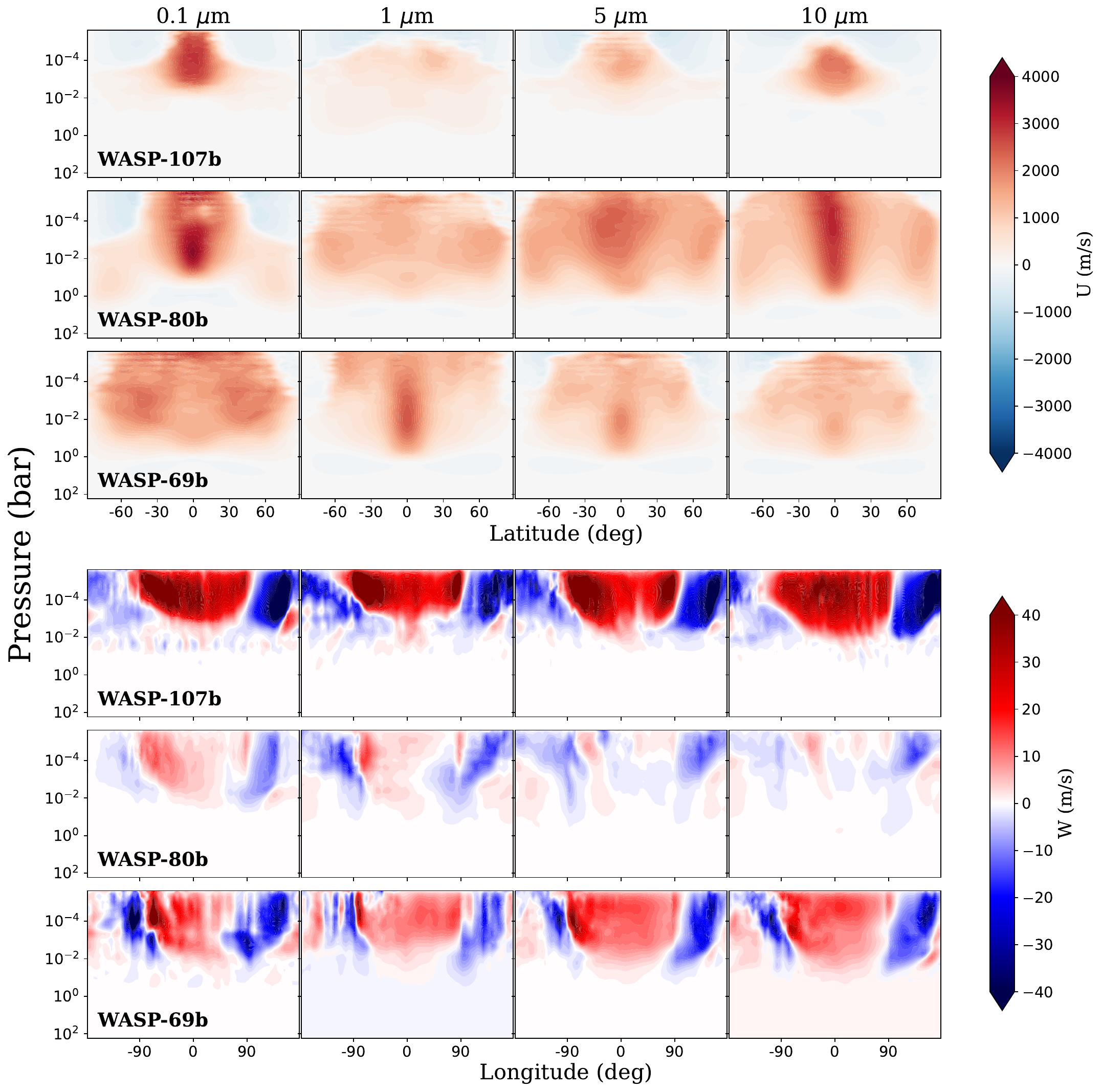}
\caption{
(Na$_2$S Models) Top: Zonal-mean zonal wind speed (latitudinal distribution of zonal component of wind (U) averaged over longitudes) for all models. Within each subplot, the x-axis shows the latitude and the y-axis the pressure. Rows 1, 2, and 3 correspond to WASP-107b, WASP-80b, and WASP-69b clouds, respectively, with each column representing a different particle size indicated at the top. Bottom: Same as top, but for longitudinal distribution of the vertical component of wind (W) averaged over latitudes. The vertical component shows the strength of upwelling on the dayside and downwelling on the nightside for different models.
}
\label{fig:uwna2s}
\end{figure*}

\begin{figure*}[!hbt]
\centering
\includegraphics[width=\textwidth]{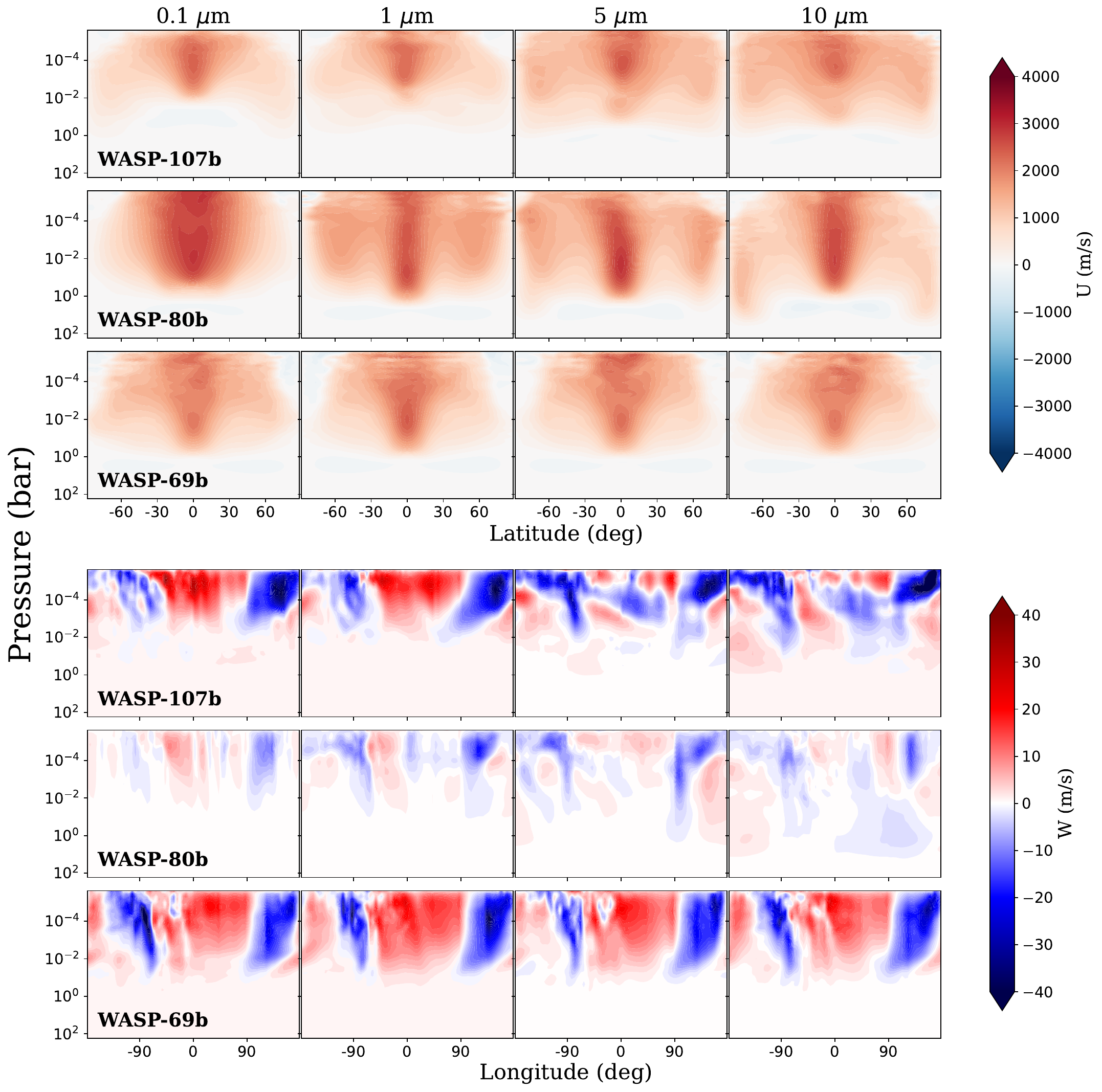}
\caption{Same as Fig. \ref{fig:uwna2s} but for KCl cloud models.
}
\label{fig:uwkcl}
\end{figure*}

\begin{figure*}[!hbt]
\centering
\includegraphics[width=\textwidth]{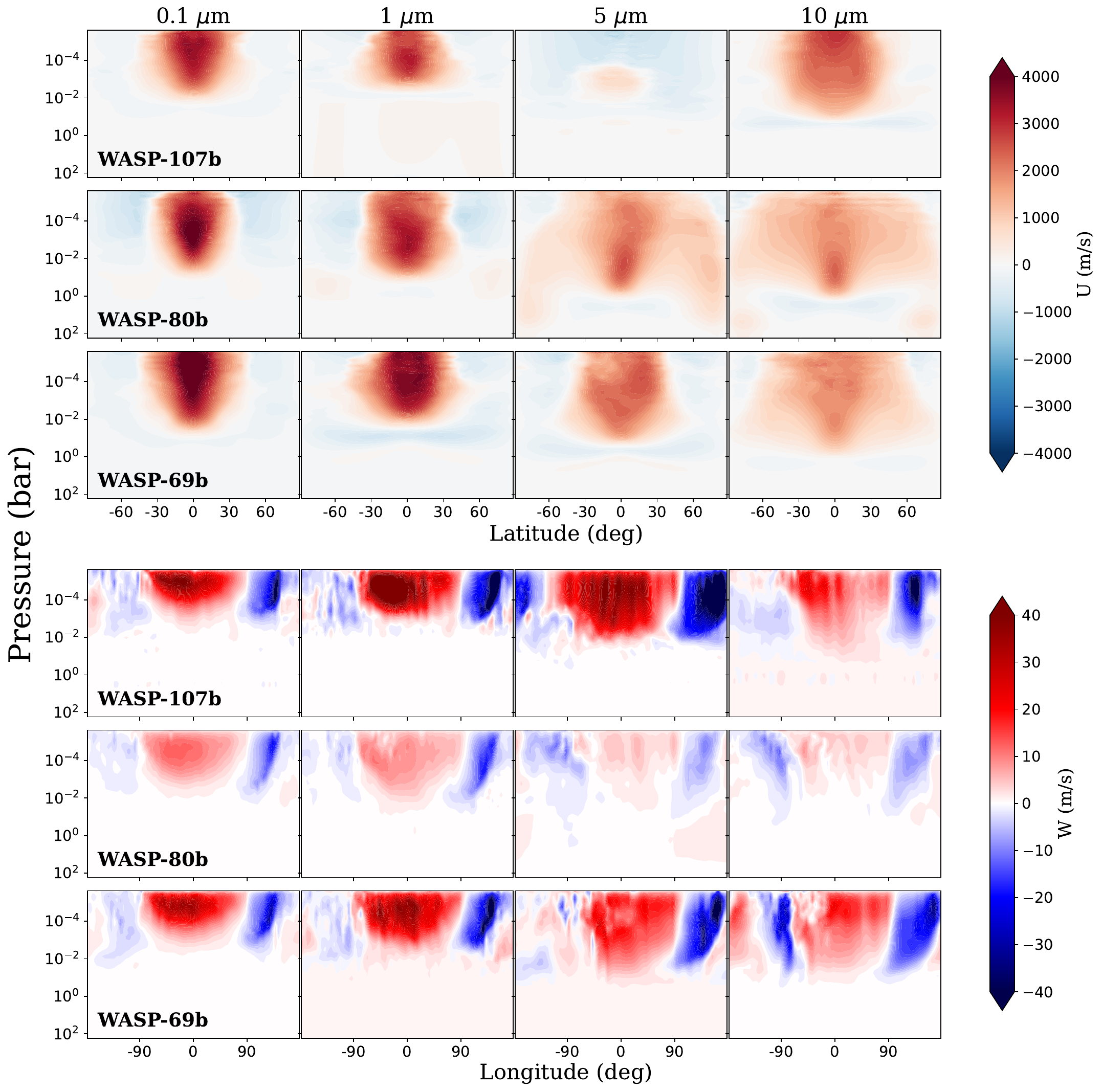}
\caption{Same as Fig. \ref{fig:uwna2s} but for MgSiO$_3$ cloud models.
}
\label{fig:uwsil}
\end{figure*}

\begin{figure*}[!hbt]
    \centering
    \includegraphics[width=\textwidth]{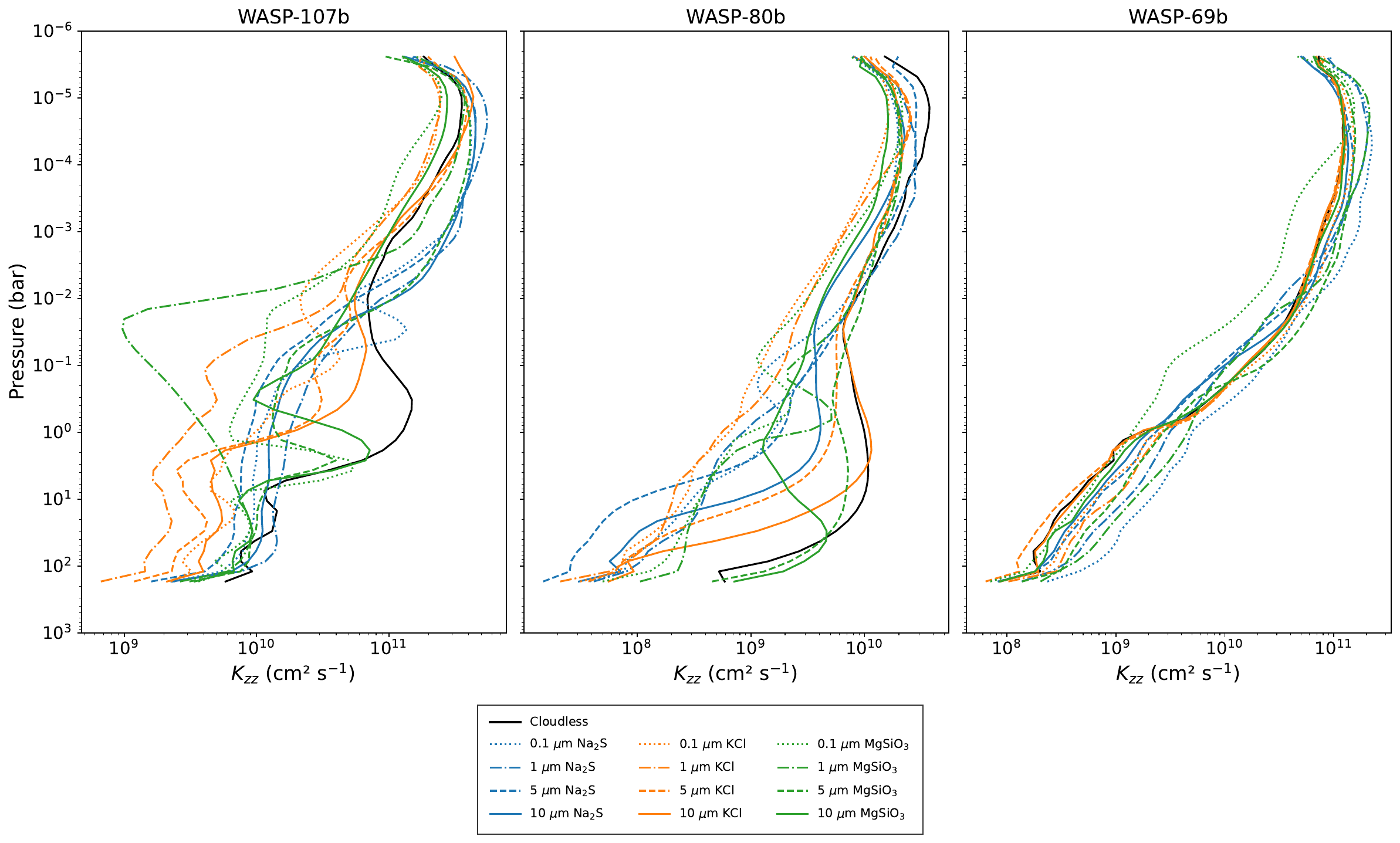}
    \caption{K$_{zz}$ profiles for WASP-107b (left), WASP-80b (middle), and WASP-69b (right) for different cloud cases. K$_{zz}$ was calculated as the root mean square of the vertical velocity times the vertical scale height.}    
    \label{fig:pkzz}
\end{figure*}

\begin{figure*}[!hbt]
\centering
\includegraphics[width=0.7\textwidth]{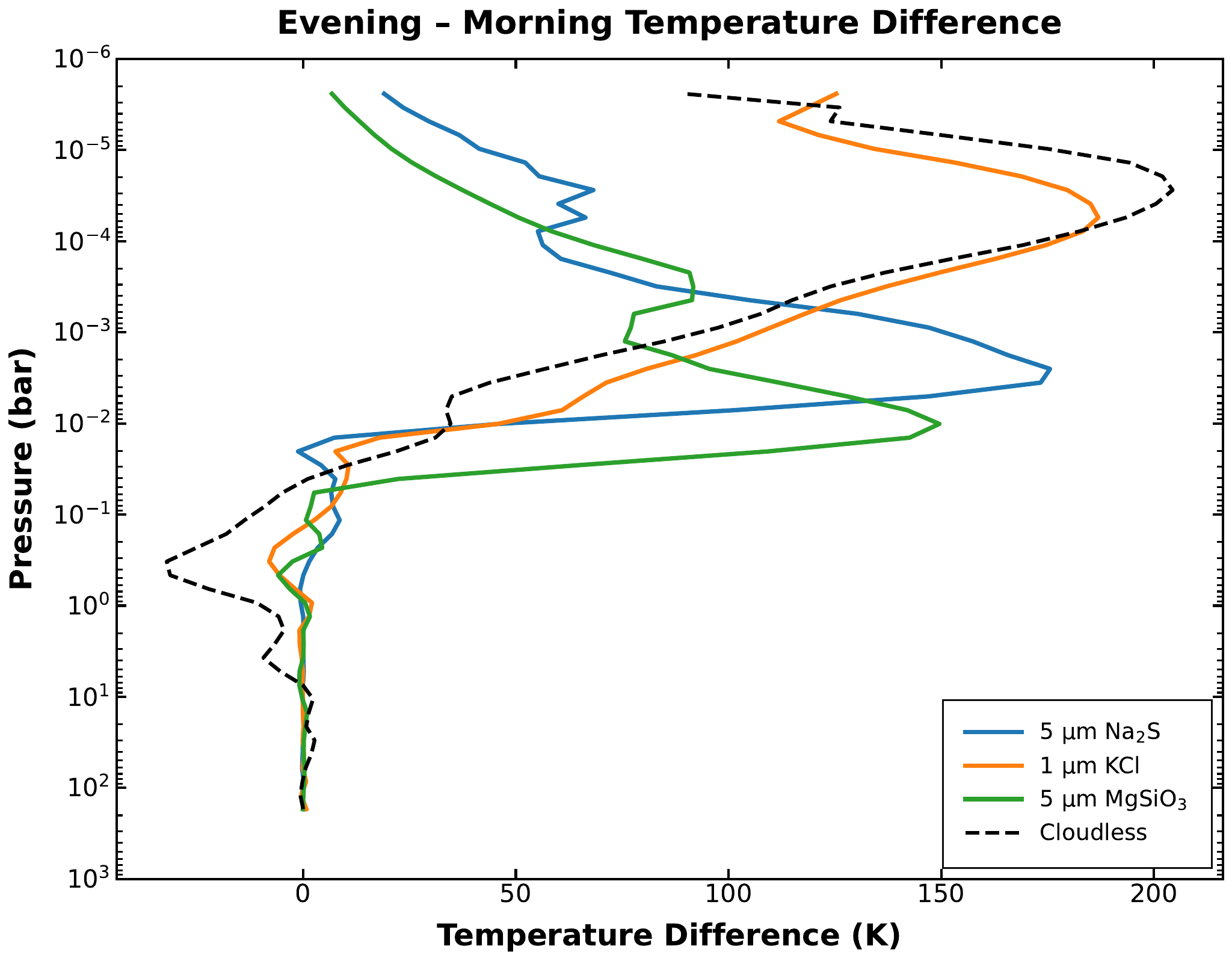}
\caption{
Vertical profiles showing temperature contrast between evening and morning terminators for different cloud compositions: 1 \micron KCl (orange), 5 \micron \mg (green), 5 \micron \nas (blue), and cloudless model (black dashed).}
\label{fig:best3}
\end{figure*}

\begin{figure*}[!hbt]
    \centering
    \includegraphics[width=\textwidth]{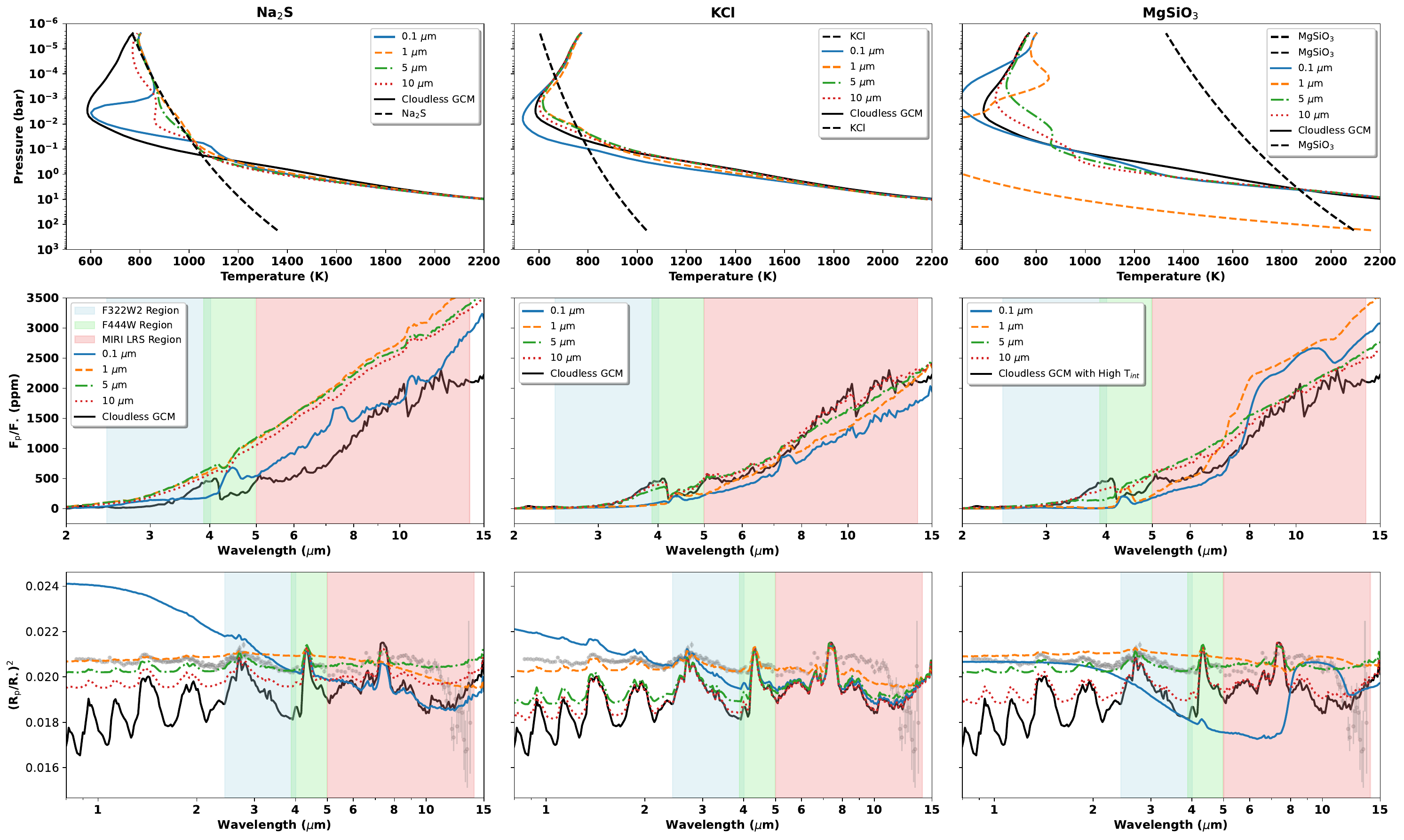}
    \caption{(WASP-107b) Top row: Dayside-averaged pressure-temperature profiles for different particle sizes along with the cloudless case (solid black line). The condensation curve for the corresponding cloud species is plotted with a black line. Middle row: Emission spectra from GCM models along with the cloudless case (solid black line). Bottom row: Transmission spectra from GCM models along with the cloudless case (solid black line) and JWST observation (gray).}
    \label{fig:tp107}
\end{figure*}

\begin{figure*}[!hbt]
    \centering
    \includegraphics[width=\textwidth]{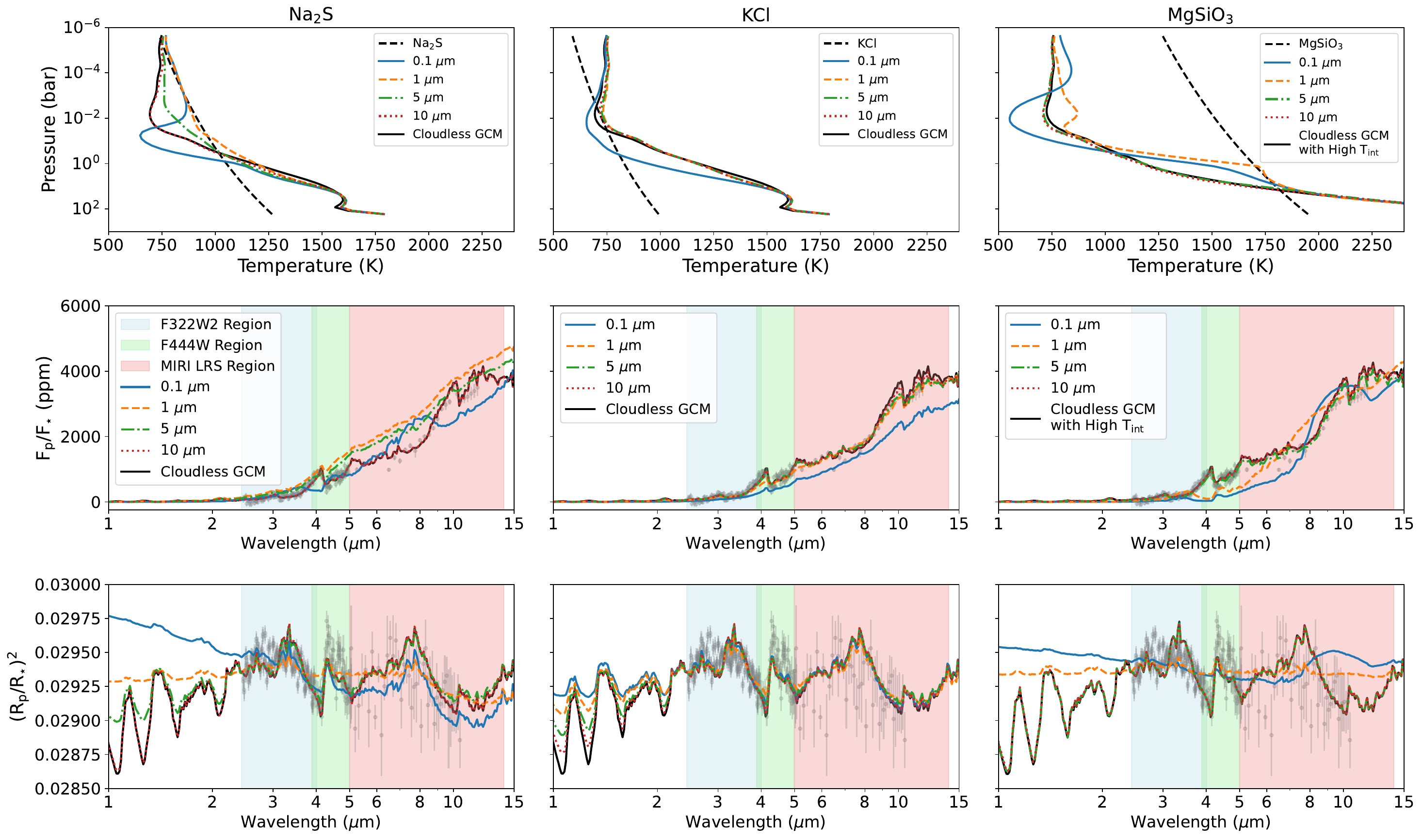}
    \caption{(WASP-80b) Top row: Dayside-averaged pressure-temperature profiles for different particle sizes along with the cloudless case (solid black line). The condensation curve for the corresponding cloud species is plotted with a black line. Middle row: Emission spectra from GCM models along with the cloudless case (solid black line) and JWST observation (gray). Bottom row: Transmission spectra from GCM models along with the cloudless case (solid black line) and JWST observation (gray). (From \citealt{mehta2026})}
    \label{fig:pspec}
\end{figure*}

\begin{figure*}[!hbt]
    \centering
    \includegraphics[width=\textwidth]{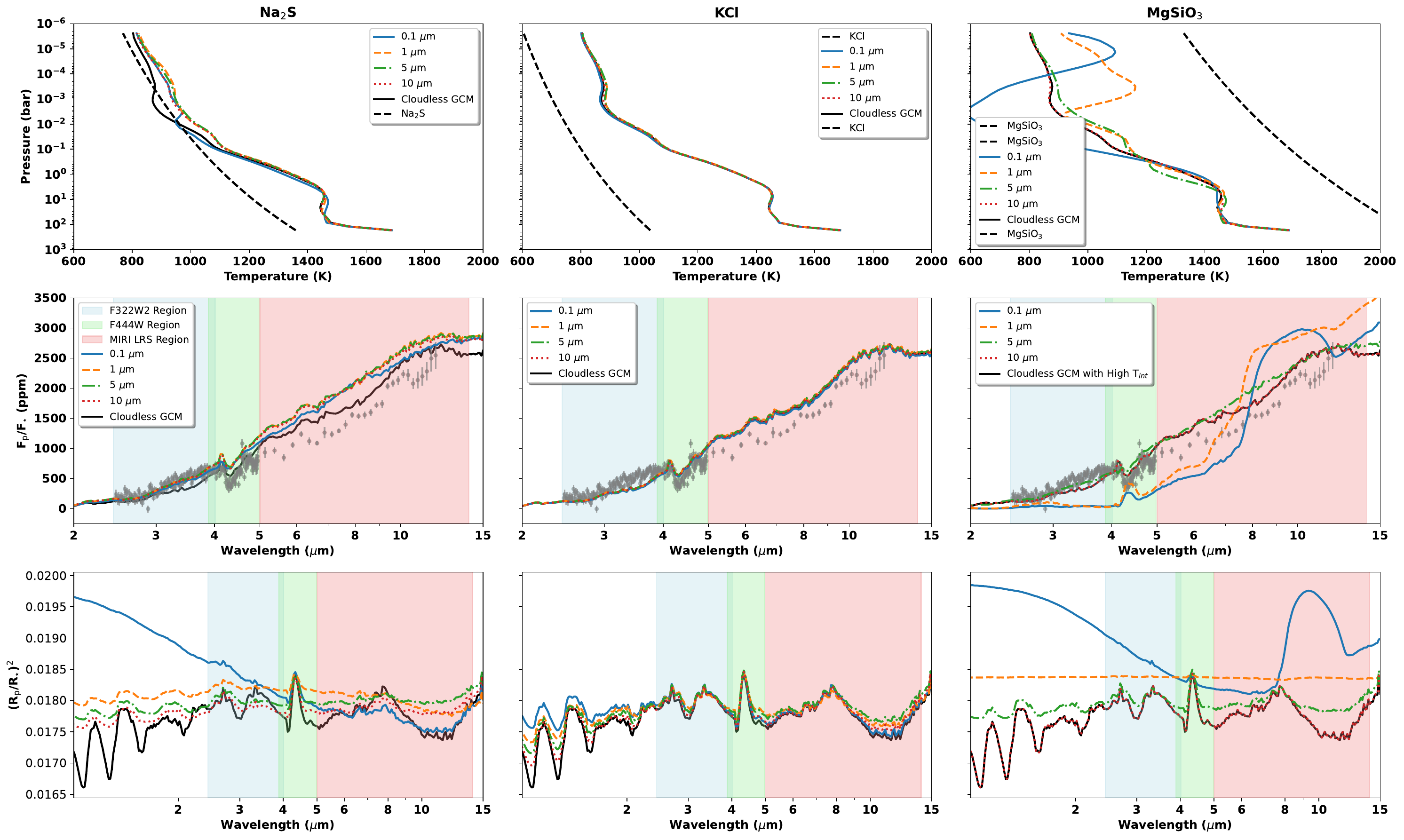}
    \caption{(WASP-69b) Top row: Dayside-averaged pressure-temperature profiles for different particle sizes along with the cloudless case (solid black line). The condensation curve for the corresponding cloud species is plotted with a black line. Middle row: Emission spectra from GCM models along with the cloudless case (solid black line) and JWST observation (gray). Bottom row: Transmission spectra from GCM models along with the cloudless case (solid black line)}
    \label{fig:tp69}
\end{figure*}

\end{appendix}

\end{document}